\pdfoutput=1
\documentclass[11pt]{article}
\usepackage[letterpaper,margin=1in]{geometry}
\usepackage[T1]{fontenc}
\usepackage{lmodern}
\usepackage{amsmath,amssymb}
\usepackage{graphicx}
\usepackage{array,booktabs,longtable}
\usepackage{pdflscape}
\usepackage[section]{placeins}
\usepackage[skip=6pt]{parskip}
\usepackage{caption}
\usepackage{microtype}
\usepackage{xurl}
\usepackage[hidelinks]{hyperref}
\title{\bfseries Animal Welfare and Policy Risk Index (AWPRI): Constructing a Cross-National Exposure Measure of Livestock and Aquatic Animal Use and Governance Capacity, 24 Countries, 2010--2022}
\author{Jason Hung\\[2pt]
\normalsize C2F Postdoctoral Fellow, Department of Sociology, Chulalongkorn University}
\date{}

\begin{document}
\maketitle

\begin{abstract}
This research paper introduces the Animal Welfare and Policy Risk Index (AWPRI), a composite exposure index covering livestock and aquatic animals used for food in 24 countries over 2010--2022 (N = 312 country-year observations, 3,744 observed values, no imputed values). The AWPRI measures animal-use exposure and governance capacity relative to the panel, not animal welfare outcomes. The AWPRI is constructed from 12 variables organised across three conceptual layers, namely Animal-use exposure (L1), Governance capacity (L2), and artificial intelligence (AI) amplification (L3). Each layer is the mean of its own variables and the composite is the mean of the three layers, so a variable sitting in the three-variable layer carries more weight than one sitting in the five-variable layer. Variables are min-max normalised within each year, so a country's value on any variable, layer, or composite suggests its position among the 24 countries that year. K-means clustering of the layer scores selects two groups (silhouette coefficient = 0.504 at k = 2, against 0.444 at k = 3 and 0.330 at k = 4). Principal component analysis (PCA) of the 12-variable 2022 cross-section requires five components to reach 90.8\% of cumulative variance. Across 46 rebuilds, the ranking is stable under alternative normalisations (rank correlation 0.982--0.994) and less stable under layer reweighting (0.914--0.995) and leave-one-variable-out (0.865--0.990). China (0.703), Vietnam (0.663), and Thailand (0.612) recorded the highest composite scores in 2022; Sweden (0.316), Kenya (0.339), and Germany (0.362) the lowest. Tested against the World Animal Protection (WAP)'s Animal Protection Index 2020, the composite returns a null on the overall grade (Spearman $\rho$ = $-$0.264, p = 0.213, n = 24) and the farm animal grade ($\rho$ = $-$0.301, p = 0.153), while L1 correlates with the overall grade only ($\rho$ = $-$0.485, p = 0.016). The composite and the WAP grades are not measuring the same quantity. The AWPRI and its interactive visualisation are publicly accessible at \url{https://awpri.aiinsocietyhub.com/}.

\end{abstract}

\section{Introduction}

Approximately 80 billion land animals are slaughtered annually within global food systems (FAO, 2023). The scale of this figure renders the institutional underinvestment in animal welfare governance both empirically significant and policy-relevant. Comparative political science and public policy scholarship have been slow to develop quantitative frameworks for cross-country welfare governance assessment. Where animal welfare is scholarly discussed, analysis is predominantly shaped by normative or legal terms (Blattner \& Bridges, 2024; H\aa{}rstad, 2024), or confined to case studies of discrete regulatory regimes (Chaney et al., 2024). There is an absence of a longitudinal, cross-country, quantitative instrument that tracks how animal welfare governance performs over time, whether those trajectories are improving or deteriorating, and whether emerging technological forces compound pre-existing governance gaps.

The rapid commercialisation of artificial intelligence (AI) in livestock production makes the discussion of technologically facilitated animal welfare regulation increasingly timely and relevant. Computer vision systems for automated lameness detection, AI-driven feed optimisation, and predictive disease modelling are now commercially deployed across major livestock-producing economies (Neethirajan, 2024; Papakonstantinou et al., 2024). In the United Kingdom, Siachos et al. (2025) evaluated a fully automated two-dimensional imaging system that uses machine learning to score the mobility of dairy cows on 11 commercial farms, and across 40,116 paired scores from the system and four trained veterinarians, the system detected lame cows with an accuracy close to that of experienced human assessors. In China, Luo et al. (2023) trained a deep learning model to detect dead hens on a commercial laying hen farm, a task otherwise done through manual inspection, and reported an average precision of 99.7\%. Market projections for the precision livestock farming (PLF) sector reach USD 19.87 billion by 2032 (DataM Intelligence, 2024). A body of literature argues that AI enables earlier detection of welfare problems and reduces reliance on invasive interventions (Schillings et al., 2021; Zhang et al., 2024). Additional literature raises substantive concerns that PLF's welfare-positive claims remain unproven at commercial scale, and that AI-driven intensification poses systemic threats to welfare in jurisdictions whose regulatory frameworks were not designed to address algorithmic accountability (Elliott \& Werkheiser, 2023; Tuyttens et al., 2022).

Existing composite measures of animal welfare governance, including the World Animal Protection (WAP)'s Animal Protection Index (2020) and H\aa{}rstad's scoping review (2024), feature legislative text at a single point in time. The WAP is an international animal protection organisation whose Animal Protection Index grades 50 countries from A (highest) to G on 10 indicators grouped under four goals. The indicators cover farm animals, animals in captivity, companion animals, animals used for draught and recreation, animals used in scientific research, and wild animals, and the indicator on the recognition of animal sentience refers to vertebrates, cephalopods, and decapod crustaceans. The 2020 edition is the second, following a 2014 edition covering the same 50 countries, and the index is designed to support the WAP's lobbying for stronger animal protection law; the WAP documentation does not report how the index is funded. H\aa{}rstad's review, written at the Norwegian University of Science and Technology and Ruralis and funded by the Research Council of Norway (grant 295161), is a scoping review of research on farm animal welfare governance published between 2000 and 2021. Neither instrument (i.e., the WAP index and H\aa{}rstad's review) tracks law enforcement dynamics, policy reform trajectories, or the compounding effect of technological adoption on governance gaps. This paper addresses these limitations through the introduction and analysis of the Animal Welfare and Policy Risk Index (AWPRI). We compare the AWPRI with the WAP index because the WAP index grades all 24 panel countries on a common scale in 2020, a year within the panel, and with H\aa{}rstad's review because the review synthesises research on farm animal welfare governance over a period (2000--2021) that overlaps the panel.

\subsection{Research Questions and Contributions}

This paper pursues three research questions. First, how can animal welfare policy risk, that is understood as the structural conditions under which welfare harms are more probable, be operationalised as a measurable, cross-country comparable composite index sensitive to the governance implications of AI adoption in agriculture? Second, what patterns of risk distribution, clustering, and temporal change emerge across 24 countries between 2010 and 2022? Third, does the composite reproduce an existing expert assessment of national animal protection law, and if it does not, which part of the index does?

\subsection{What the Index Measures and What It Does Not}

Five terms are used throughout this paper. Animal welfare is the state of an individual animal as that animal experiences its own conditions (Browning, 2022). A welfare harm is a specified event that worsens that state, such as confinement that prevents an animal from turning around, transport beyond a duration the animal can tolerate without distress, or slaughter without effective stunning. Animal protection is the body of law and institutional practice meant to prevent those harms. Animal rights is a normative position about the moral standing of animals. Governance capacity is the institutional condition under which animal protection law can be enforced. Policy risk, as used here, is the structural condition under which welfare harms are more probable. The AWPRI measures animal-use exposure, governance capacity, and the research composition behind agricultural automation. Our AWPRI does not measure animal welfare, and no score within our index should be read as a statement about how animals in a country fare.

The animal populations in scope are terrestrial farmed animals slaughtered for food and aquatic animals entering the food supply, whether farmed or wild-caught, because the FAOSTAT Food Balance Sheets do not separate farmed from wild-caught aquatic animals. Companion animals, laboratory animals, and wild animals outside the food supply sit outside the panel, because none of the four Layer 1 series distinguishes these animals and a score would otherwise contain a quantity the sources do not report. The sectors in scope are livestock production, aquaculture, capture fisheries, and the national research base that supplies automation to livestock production and aquaculture. Table A1 (Appendix) states, for each of the 12 variables, the welfare harm the variable is meant to stand for and whether the pathway from indicator to harm is (1) direct, (2) indirect, or (3) assumed. Three variables carry an assumed pathway and all three sit in Layer 3.

\section{Literature Review}

\subsection{Animal Welfare Governance and Composite Indices}

Composite indices are well-established instruments for cross-country governance comparison. The Human Development Index (UNDP, 1990), the Environmental Performance Index (Wolf et al., 2022), and the Global Peace Index (IEP, 2024) demonstrate that multidimensional governance circumstances can be reduced to measurable scores while retaining policy interpretability. In the animal welfare domain, Browning (2022) argues explicitly that multidimensional welfare measurement frameworks can support policy analysis. However, to date, no composite index applies such a framework to the intersection of animal welfare governance and AI adoption.

The WAP's Animal Protection Index (2020) rates 50 countries on legislative capacity at a single point in time. H\aa{}rstad's scoping review of farm animal welfare governance (2024) similarly prioritises legislative text and political drivers. H\aa{}rstad finds that policy change is neither linear nor easily predictable. Neither instrument takes into account the enforcement dynamics, temporal trajectories, or the risk introduced by AI-driven agricultural intensification. The Animal Law Foundation (2024) has documented that fewer than 2.5\% of farms in England were inspected in 2024, with 19\% of inspected farms found in breach of welfare laws and fewer than 1\% of violations resulting in prosecution. This enforcement gap illustrates the dimension that static legislative indices fail to highlight.

\subsection{AI and PLF}

Tuyttens et al. (2022) identify 12 welfare threats specific to PLF adoption, including the displacement of human observation by algorithms, the intensification of stock enabled by automated monitoring, and the commercial incentives to use AI for productivity maximisation over welfare improvement. These concerns are amplified in jurisdictions with weak baseline welfare legislation, in which AI adoption can accelerate production intensification without triggering comparable regulatory responses (Parlasca et al., 2023). Elliott and Werkheiser (2023) argue that existing PLF transparency frameworks remain conceptually underdeveloped, and that most AI agricultural systems operate without welfare-specific accountability mechanisms.

\subsection{Panel Data Approaches to Governance Measurement}

Fixed-effects panel regression and event-study designs are standard approaches to exploiting longitudinal cross-country differences in governance indices (Angrist \& Pischke, 2009; Callaway \& Sant'Anna, 2021). Where the treatment is itself one of the variables that contributes to the outcome, the estimate is tied to the outcome and the research design measures the split of countries into groups and not its effect.

\section{Methods}

\subsection{Country and Time Period}

The AWPRI panel covers 24 countries across six World Bank regions (World Bank, n.d.) over 13 years (2010--2022), giving 312 country-year observations. Under this classification, Canada and the United States are grouped within North America; the Middle East and North Africa region is absent from the panel because no country meeting the coverage criterion falls within that region, not by regional design. Our panel dataset begins with 2010 data because the FAOSTAT Food Balance Sheets have available data from 2010. FAO revised the Food Balance Sheet methodology in 2020 and re-estimated only back to 2010. While we initially planned to include Japan, yet such a country's data are present in the FAOSTAT's production data and absent from its Food Balance Sheets. Therefore, we dropped Japan when compiling our panel dataset. No country is retained on regional or substantive grounds where the coverage criterion is not met, and no country-year value is imputed, interpolated, carried forward, or reconstructed. The panel contains 3,744 values and all 3,744 are observed. Table A2 (Appendix) reports the number of observed values for every country. The 24 countries are Argentina, Australia, Brazil, Canada, China, Denmark, France, Germany, India, Italy, Kenya, Mexico, the Netherlands, New Zealand, Nigeria, Poland, South Africa, South Korea, Spain, Sweden, Thailand, the United Kingdom, the United States, and Vietnam. Three major livestock-producing economies, namely Indonesia, Pakistan, and Ethiopia, were not part of the country set assembled for this panel; unlike Japan, they were never carried through data collection, so no coverage figure is reported for them here. Their inclusion is a priority for the scale-up phase of the AWPRI project. All analyses reported in this paper use the \textit{panel\_\allowbreak awpri\_\allowbreak normalized.csv} dataset, which is the same file that generates every figure on the AWPRI site, and it is published with the codebook at \url{https://awpri.aiinsocietyhub.com/}.

\subsection{Variable Selection and Operationalisation}

A total of 12 variables are assigned across three conceptual layers. Layer 1 (L1, Animal-use exposure) measures how many animals a country's food system uses and of what kind, through (1) farmed animals slaughtered per capita, (2) the aquatic animal share of flesh supply, (3) meat supply per capita, and (4) the animal share of protein supply. Layer 2 (L2, Governance capacity) measures the stock of binding welfare law and the institutional conditions under which welfare law is enforced, through (5) the cumulative stock of binding animal welfare instruments, (6) rule of law, (7) the five-year flow of new binding animal welfare instruments, (8) civic space, and (9) civil liberties. Layer 3 (L3, AI amplification) measures how national research effort divides between automating animal agriculture and studying animal welfare, through (10) AI research intensity, (11) precision agriculture research intensity, and (12) the animal welfare research base. All 12 variables are coded so that a higher value means more animals exposed, weaker welfare law, worse institutional conditions for enforcing welfare law, or research effort tilted further towards automating animal agriculture. Data are drawn from FAOSTAT (Crops and Livestock Products, Food Balance Sheets, Macro-Statistics and Population; FAO, 2026), FAOLEX (FAO, 2025), V-Dem v15 (V-Dem Institute, 2025), and OpenAlex (Priem et al., 2022). The FAOSTAT bulk files were downloaded on 4 August 2026 and come from the releases of 23 December 2025 (Crops and Livestock Products), 14 October 2025 (Food Balance Sheets), 27 March 2026 (Macro-Statistics), and 4 March 2026 (Population), and FAOLEX and OpenAlex were retrieved on the same date. Table A1 (Appendix) gives the query behind every series, the welfare harm the variable stands for, and whether the pathway from indicator to harm is direct, indirect, or assumed. Three of the 12 carry an assumed pathway and all three sit in Layer 3. Layer 3 is built to stand for the compounding risk associated with the governance-technology asymmetry described in Section 2.2, and the pathway from a national research base to a welfare harm is assumed and not established. Section 4.8 reports what the composite looks like when Layer 3 is removed. In our index, we do not directly measure enforcement intensity. National farm inspection rates and prosecution rates are not published longitudinally across 24 countries for 13 years (2010-22), and Section 6 states such a circumstance as a limitation and not as a gap that the L2 variables close.

Three properties of the sources depend on how the scores should be read. First, FAOSTAT counts slaughter within national boundaries irrespective of where the animals were born or raised, so VAR\_01 attributes an animal to the country in which the animal is slaughtered, whereas Food Balance Sheet supply is production plus imports minus exports, net of stock changes, so VAR\_05 and VAR\_07 attribute meat and protein to the country of supply. L1 therefore combines animals slaughtered within a country's borders with animal products supplied to the country's population, wherever those animals were raised. Second, FAOLEX records the enactment of an instrument and not its enforcement, so the two legislation variables measure how much binding welfare law a country has enacted and not whether that law is applied. The denominator of both legislation variables is the number of records indexed for the country at retrieval, held fixed across years, and ranges from 566 (South Korea) to 6,056 (the United Kingdom), including 2,626 of the 62,374 records across the 24 countries that are dated after 2022. EU instruments are attributed to each of the eight member states and to the United Kingdom up to 2020. Third, OpenAlex works are counted in full, so a work with authors in more than one country counts once for each of those countries, in both the numerator and the national total. Works are identified by a search of titles and abstracts for the terms listed in Table A1 (Appendix). The three Layer 3 variables therefore measure research output and not the use of AI on farms, and the link between the two is the assumed pathway stated in Table A1. A search for works that contain both an AI term and a welfare term returned between 0 and 7 United Kingdom works a year, too few for an annual series, so AI research and welfare research are measured as separate series.

\subsection{AWPRI Construction}

All 12 variables are normalised to [0, 1] using min-max normalisation within each year, so that a score states where a country sits among the 24 in that year and its trajectory over time is relative to the counterpart of the panel. Before normalisation, a variable is transformed to log(1 + x) where the pooled skewness of the variable exceeds +1 and the transformation brings the absolute skewness below 1. Three variables meet this rule, namely AI research intensity (skewness 1.98 before and 0.93 after), precision agriculture research intensity (2.41 and 0.51), and the animal welfare research base (1.45 and 0.17). Civic space (2.06 and 1.72) and civil liberties (1.50 and 1.29) are not transformed, because the skewness of both stays above 1 after transformation. After normalisation, the two legislation variables and the animal welfare research base are inverted (1 $-$ v), so that more welfare law and a larger welfare research base each lower the score, and the three V-Dem variables are inverted at source (Table A1, Appendix). The welfare research base is reversed because Layer 3 reads the two automation variables against the national expertise that can evaluate and contest new husbandry technology, so a larger welfare research base is coded as lower risk. Each layer score is the unweighted mean of its own constituent variables:

\[
L_{1it} = \frac{1}{4} \sum_{k \in \mathcal{K}_1} v_{kit}
\]

\[
L_{2it} = \frac{1}{5} \sum_{k \in \mathcal{K}_2} v_{kit}
\]

\[
L_{3it} = \frac{1}{3} \sum_{k \in \mathcal{K}_3} v_{kit}
\]

where $\mathcal{K}_1$, $\mathcal{K}_2$, and $\mathcal{K}_3$ denote the four-variable, five-variable, and three-variable indicator sets for Layer 1, Layer 2, and Layer 3 respectively, as defined in Table A1 (Appendix), and where $v_{kit}$ denotes the normalised value of variable $k$ for country $i$ in year $t$.

The composite AWPRI score is the unweighted mean of the three layer scores:

\[
\mathit{AWPRI}_{it} = \left(L_{1it} + L_{2it} + L_{3it}\right) / 3
\]

Because the layers hold four, five, and three variables and each layer contributes one third of the composite, the variables are not equally weighted against one another. A Layer 3 variable carries 1/9 (= 1/3 x 1/3) of the composite, a Layer 1 variable 1/12 (= 1/4 x 1/3), and a Layer 2 variable 1/15 (= 1/5 x 1/3). Equal weighting at the layer level is applied following Organisation for Economic Co-operation and Development (OECD) and Joint Research Centre (JRC) recommendations for composite indicators when no strong prior evidence exists for differential weighting across dimensions (OECD \& JRC, 2008). The consequences of our equal weighting decision are reported in Section 4.8.

\subsection{Cluster Analysis}

Our initial analysis defined a four-tier risk typology using score-based thresholds (Critical $\geq$ 0.55, High 0.45--0.55, Moderate 0.35--0.45, Low < 0.35) and then reported the associated k-means results. Those thresholds were set by us and not by the data. However, we decided to withdraw such a four-tier risk typology. Cluster structure is instead determined by k-means on the 2022 cross-section of the three layer scores, with k chosen on the silhouette coefficient across k = 2 to k = 7, and we report the associated Calinski--Harabasz and Davies--Bouldin indices. Whatever k the silhouette selects is the partition this research paper uses, and no tier labels are assigned to the resulting groups beyond which group holds the higher mean composite score.

\subsection{Trend Projection}

An initial analysis projected country trajectories to 2030 using AutoRegressive Integrated Moving Average (ARIMA) models with order selection by Akaike Information Criterion minimisation. On a 13-year series that specification is selecting an order from 13 points. In that initial analysis the specification returned no autocorrelation structure being detected for most countries, which means the projection was the country mean carried forward while being presented as a model. We replace the ARIMA specification with an ordinary least squares (OLS) linear trend on year, fitted separately for each country over 2010--2022 and extrapolated to 2030 with the prediction interval that follows from the model fit. The OLS trend extrapolation assumes the 2010--2022 slope continues, and Table S1 (Supplementary Material) reports the slope p value alongside every projection so that a reader can see which of them relies on a slope distinguishable from zero.

\subsection{Statistical Analysis}

We conducted a total of six complementary inferential analyses.

First, Wilcoxon signed-rank tests are used to compare the three layer scores against one another, both across the full panel (N = 312) and in the 2022 cross-section (n = 24). The signed-rank test is preferred over the parametric t-test given the bounded, non-normal distribution of normalised layer scores. A layer holding a higher mean than another is a statement about where the panel is positioned on that layer relative to itself, and not a statement that one kind of risk is larger than another in any absolute term. Second, a Spearman rank correlation matrix is computed for the 12 constituent variables on the 2022 cross-section to assess construct validity and detect potential multicollinearity in the index structure. Third, a Kruskal--Wallis test and a Mann--Whitney U test are applied to the k-means partition selected in Section 3.4, to test whether composite scores differ across the groups the algorithm returns. Fourth, our index is tested against an external assessment of national animal protection law. Composite, layer, and single-variable scores for 2020, the year of the WAP assessment, are correlated against the WAP's Animal Protection Index 2020 overall grade and its farm animal grade, which is the grade on the indicator for protecting animals used in farming (WAP, 2020), both as raw Spearman correlations and partial correlations net of log gross domestic product per capita. The WAP grades are ordinal and are scored so that a numerically higher value denotes a better grade, which means a working index should return a negative correlation.

The correlation is Spearman's $\rho$:

\[
\rho = 1 - \frac{6 \sum d_i^2}{n\left(n^2 - 1\right)}
\]

where $d_i$ is the difference between the rank of country $i$ on the AWPRI and its rank on the WAP grade, and $n = 24$. Ties in the WAP grades are handled by mid-ranks.

Fifth, an event study estimates whether a dated, binding instrument moved the index in the countries it binds. Our initial analysis reported a difference-in-differences estimate in which countries were split on \textit{ai\_\allowbreak governance\_\allowbreak risk}, a variable that was one of five constituents of L3 and therefore one fifteenth (= 1/5 x 1/3) of the composite it was being used to explain. That design estimated the split, which is what produced a coefficient of 0.200 on L3 with a standard error of 0.000 and t = 2515.9. Group membership was also fixed by a 2019 assessment while treatment was dated 2017, so status was determined partly by what happened during the treatment period. The variable that defined the treatment is one of the three withdrawn variables listed in Table A1 (Appendix). Our difference-in-differences design is withdrawn fully. \textit{Regulation (EU) 2017/625} on official controls (European Parliament \& Council of the European Union, 2017), in application from 14 December 2019, replaces the difference-in-differences design. The regulation is external to the index. This regulation binds a defined set of countries and governs~how member states inspect and enforce, animal welfare included. The treated group is the eight European Union (EU) member states in the panel (Denmark, France, Germany, Italy, the Netherlands, Poland, Spain, and Sweden) plus the United Kingdom, which was bound by the EU law on 14 December 2019; the control group is the remaining 15 countries. The event-study estimating equation is:

\[
Y_{it} = \alpha + \sum_{k \neq -1} \beta_k D_{kit} + \gamma_i + \delta_t + \varepsilon_{it}
\]

where $Y_{it}$ is the legislation-purged composite for country $i$ in year $t$, $D_{kit}$ is an indicator for country $i$ being $k$ years from the application date, $\gamma_i$ and $\delta_t$ denote country and year fixed effects respectively, and $k = -1$ is omitted as the reference. Event time runs from $k = -9$ (2010) to $k = +3$ (2022). Standard errors are clustered by country. Two variants are estimated alongside the main specification, one dropping the United Kingdom and one dating the event to 2020, and the coefficients of all three specifications are reported in Section 4.6 whether or not the coefficients are statistically significant.

Sixth, a principal component analysis (PCA) of the standardised 12-variable 2022 cross-section is conducted to identify the latent dimensional structure of the index and to determine whether the three layers are reducible to a single component. The robustness of our construction is presented in Section 3.7.

\subsection{Robustness of the Construction}

Three of our construction choices in Section 3.3 are arbitrary, namely (1) min-max normalisation in place of standardisation, (2) within-year bounds set by whichever countries happen to be in the panel, and (3) equal weights at the layer level. The construction is also tested for whether any single variable is carrying the result. Our index is rebuilt under 46 alternative specifications, and each rebuild is compared to the index as built on two quantities, namely the Spearman rank correlation of the two country orderings and the largest individual rank move. Both quantities are reported. The 46 rebuilds are three alternative normalisations, four alternative layer weightings, one removal of each of the 12 variables, one removal of each of the three layers, and one removal of each of the 24 countries.

\section{Results}

\subsection{Descriptive Statistics}

Table 1 presents summary statistics for the AWPRI composite score and its three constituent layer scores across the full panel (N = 312) and in the 2022 cross-section (n = 24). The AWPRI has a full-panel mean of 0.440 (SD = 0.101) and is positively skewed (skewness = 0.79). The positive skewness indicates a right-tailed distribution in which the majority of country-year observations cluster at moderate-to-low values, with a sparse concentration of high scores at the upper extreme. L1 records the highest full-panel mean (0.532, SD = 0.149) and L2 the lowest (0.351, SD = 0.233). The mean within-country standard deviation of L1 (0.020) and L2 (0.023), against L3 (0.088), indicates that a country's animal-use profile and governance capacity change slowly across the panel while its research composition behind agricultural automation changes quickly.

\begin{table}[!htbp]
\centering
\caption{Summary Statistics, AWPRI and Layer Scores, Full Panel (2010--2022, N = 312) and 2022 Cross-Section (n = 24)}
\label{tab:1}
\small
\begin{tabular*}{\linewidth}{@{\extracolsep{\fill}}llllllll@{}}
\toprule
\textbf{Variable} & \textbf{N} & \textbf{Mean} & \textbf{SD} & \textbf{Min} & \textbf{Median} & \textbf{Max} & \textbf{Skewness} \\
\midrule
AWPRI & 312 & 0.440 & 0.101 & 0.227 & 0.432 & 0.745 & 0.79 \\
L1, Animal-use exposure & 312 & 0.532 & 0.149 & 0.135 & 0.556 & 0.779 & $-$0.87 \\
L2, Governance capacity & 312 & 0.351 & 0.233 & 0.006 & 0.289 & 1.000 & 1.13 \\
L3, AI amplification & 312 & 0.436 & 0.154 & 0.059 & 0.415 & 0.881 & 0.39 \\
AWPRI (2022) & 24 & 0.464 & 0.099 & 0.316 & 0.445 & 0.703 & 0.88 \\
L1, Animal-use exposure (2022) & 24 & 0.554 & 0.152 & 0.158 & 0.567 & 0.757 & $-$1.17 \\
L2, Governance capacity (2022) & 24 & 0.362 & 0.239 & 0.017 & 0.291 & 1.000 & 1.16 \\
L3, AI amplification (2022) & 24 & 0.475 & 0.139 & 0.215 & 0.457 & 0.881 & 0.86 \\
\bottomrule
\end{tabular*}
\par\smallskip
{\footnotesize\raggedright \textit{Note. The first four rows report statistics for the full panel (2010--2022, N = 312), and the last four rows, marked (2022), report statistics for the 2022 cross-section only (n = 24). AWPRI is the composite score, L1 is Animal-use exposure, L2 is Governance capacity, and L3 is AI amplification. Each of the 12 variables is min-max normalised within each year, so the variable runs from 0 (the lowest value among the 24 countries in that year) to 1 (the highest), and layer and composite scores are means of these values. Higher values indicate greater exposure, weaker governance capacity, or research effort tilted further towards automating animal agriculture.}\par}
\end{table}

In the 2022 cross-section (n = 24), the sample mean AWPRI is 0.464 (SD = 0.099), and Figure 1 presents the 2022 AWPRI ranking of the 24 countries. L1 records the highest mean at 0.554 (SD = 0.152), exceeding L3 (mean = 0.475, SD = 0.139) and L2 (mean = 0.362, SD = 0.239). The maximum L3 score is recorded by India (0.881); the minimum by New Zealand (0.215). Wilcoxon signed-rank tests across the full panel show that L1 exceeds L2 (W = 11,290, p < 0.001), that L3 exceeds L2 (W = 13,123, p < 0.001), and that L1 exceeds L3 (W = 12,985, p < 0.001). In the 2022 cross-section the L1 and L2 difference holds (W = 60, p = 0.009) and the L3 and L2 difference holds (W = 69, p = 0.019), while the L3 and L1 difference does not reach conventional level of statistical significance (W = 86, p = 0.069).

\begin{figure}[!htbp]
\centering
\includegraphics[width=0.919\linewidth]{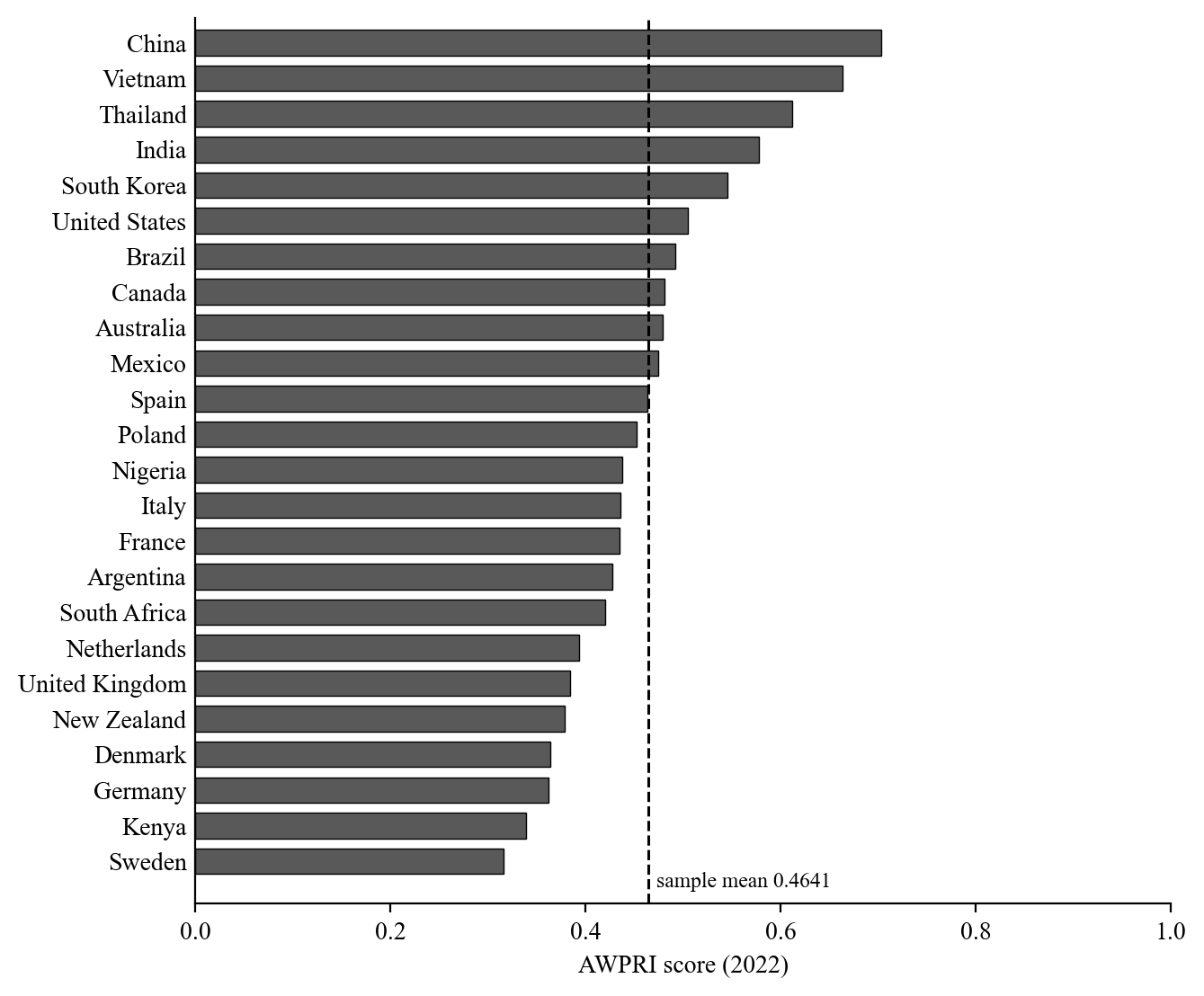}
\caption{AWPRI Rankings by Country, 2022. Countries ordered by AWPRI score (descending). Dashed line = sample mean (0.4641).}
\label{fig:1}
\end{figure}

\subsection{Spearman Correlation Structure}

Figure 2 presents pairwise Spearman rank correlations among the 12 constituent variables in the 2022 cross-section. We see several high correlations in Figure 2, most notably between \textit{rule\_\allowbreak of\_\allowbreak law\_\allowbreak risk} and \textit{civil\_\allowbreak liberties\_\allowbreak risk} ($\rho$ = 0.94), and between \textit{welfare\_\allowbreak legislation\_\allowbreak stock\_\allowbreak risk} and \textit{welfare\_\allowbreak legislation\_\allowbreak flow\_\allowbreak risk} ($\rho$ = 0.91). Both pairs are within L2 and both are retained on theoretical grounds. Here, \textit{rule\_\allowbreak of\_\allowbreak law\_\allowbreak risk} shows formal institutional constraints on arbitrary state action, whereas \textit{civil\_\allowbreak liberties\_\allowbreak risk} features the practical exercise of individual freedoms, representing separable dimensions of the governance environment. The highest cross-layer correlation is between \textit{animal\_\allowbreak protein\_\allowbreak share} and \textit{rule\_\allowbreak of\_\allowbreak law\_\allowbreak risk} ($\rho$ = $-$0.78), which states that countries drawing more of their protein from animals tend to be countries with stronger rule of law, and it is negative in coefficient value because the two variables are risk-coded in opposite directions with respect to income. The highest correlation involving L3 is between \textit{aquatic\_\allowbreak share\_\allowbreak of\_\allowbreak flesh} and \textit{ai\_\allowbreak research\_\allowbreak intensity} ($\rho$ = 0.66). No pair exceeds 0.95 in coefficient value (= |$\rho$|), implying that no variable is redundant with another under any conventional threshold. Figure 2 presents the full 12 $\times$ 12 Spearman correlation heatmap.

\begin{figure}[!htbp]
\centering
\includegraphics[width=0.983\linewidth]{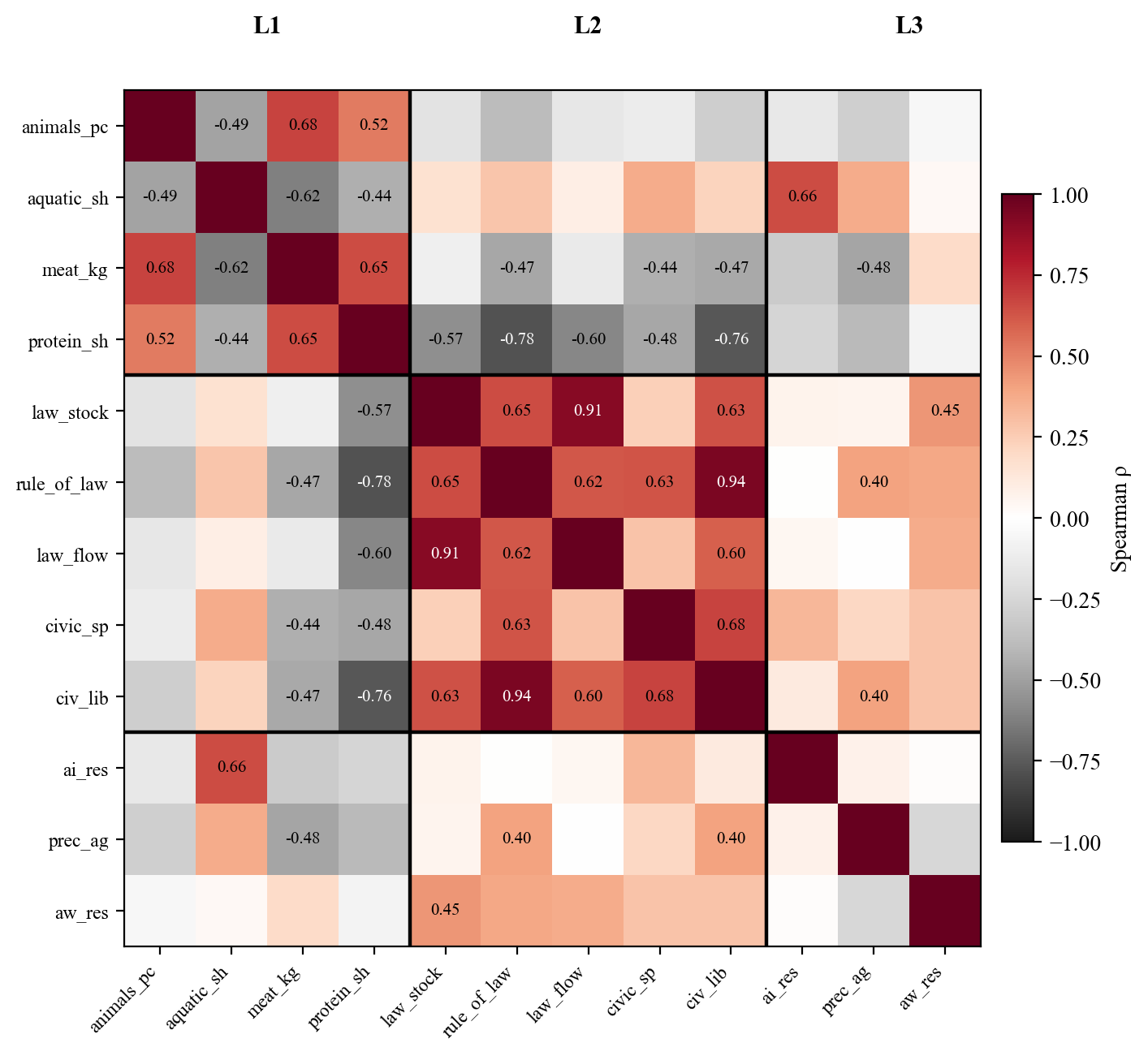}
\caption{Spearman Rank Correlation Matrix, 12 Constituent Variables, 2022 Cross-Section. Bold horizontal and vertical lines delineate L1, L2, and L3 boundaries. Values displayed where |$\rho$| $\geq$ 0.40.}
\label{fig:2}
\end{figure}

\subsection{Cross-Country AWPRI Scores, 2022}

Table 2 presents the AWPRI composite scores and layer decompositions for all 24 countries as of 2022, in the order shown in Figure 1. In 2022, China recorded the highest AWPRI score (0.703), driven by the highest L2 score in the sample (0.999). China holds the weakest position in the panel on governance capacity and the lowest rule of law, civic space, and civil liberties scores among the 24. We do not read this as a deteriorating reform trajectory. L2 measures a country's current level of governance capacity, not the rate at which that capacity is changing, and China's L2 score is close to the ceiling in every year of the panel. Vietnam (0.663) and Thailand (0.612) record the second and third highest composite scores, with Vietnam recording an L3 score of 0.657 and Thailand 0.558, both above the sample mean (0.475). At the lower end, Sweden records the lowest AWPRI score (0.316), followed by Kenya (0.339) and Germany (0.362). Kenya sitting second from bottom is worth stating plainly, because Kenya reaches that position on the lowest L1 score in the panel (0.158) and not on governance capacity, where the country sits in the weaker half (L2 = 0.452). A low composite score here means low animal-use exposure relative to the panel, and it does not mean that animals in Kenya are well protected.

Figure 3 presents the layer score decomposition across all 24 countries. The pattern that the initial analysis reported, in which L3 exceeded L1 and L2 for most of the sample, does not remain valid in our rebuild. In 2022, L1 is the highest of the three layers for 16 of the 24 countries, and the eight exceptions are China, India, Italy, Kenya, Nigeria, South Africa, Thailand, and Vietnam. High-income countries in this panel carry high animal-use exposure and strong governance capacity at the same time, and the composite is the average of the two. This is a property of the construction and is stated here so that the country ordering in Table 2 is read correctly.

\begin{figure}[!htbp]
\centering
\includegraphics[width=0.960\linewidth]{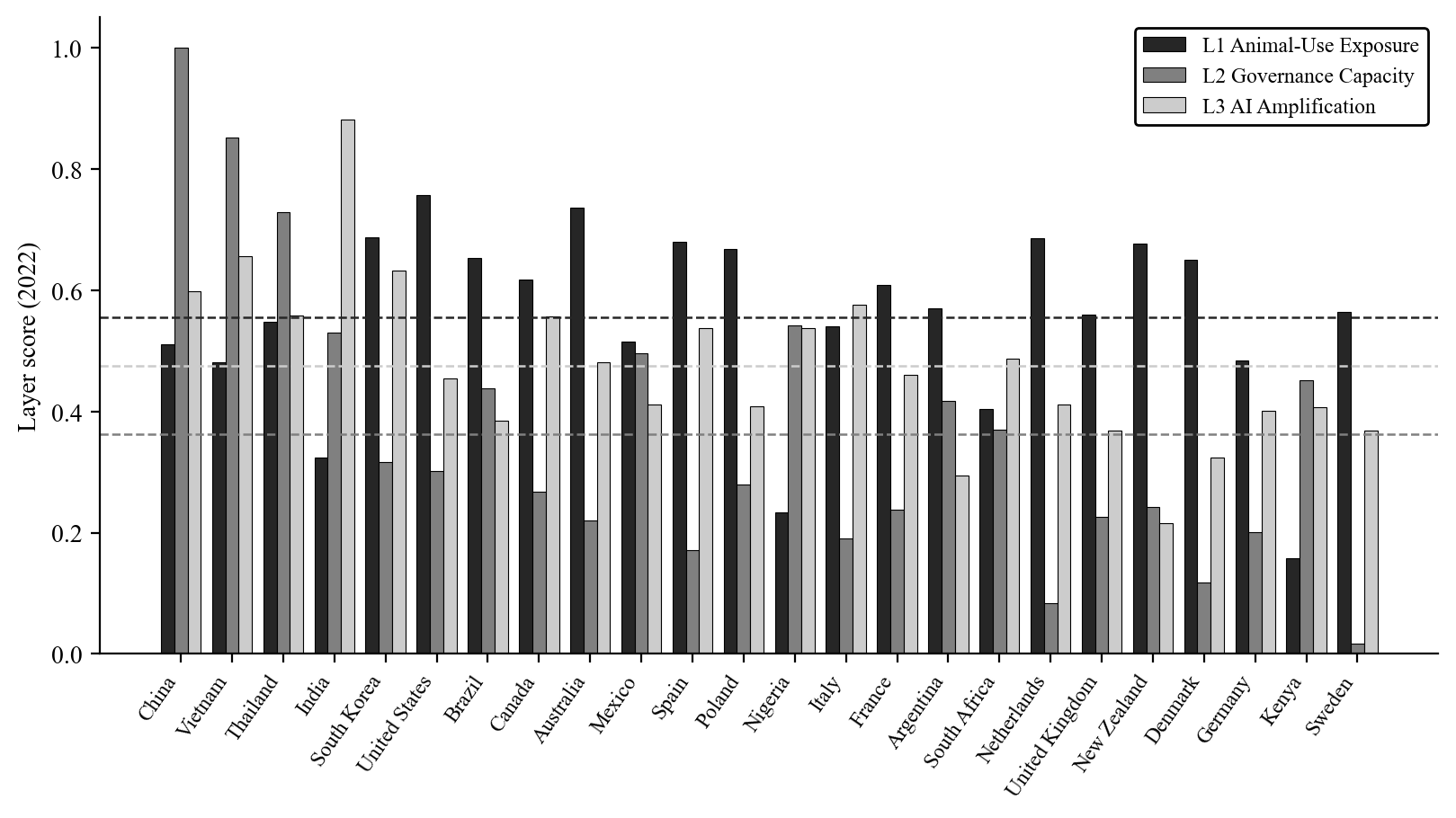}
\caption{Layer Score Decomposition by Country, 2022. Countries ordered by AWPRI score (descending). Dashed horizontal lines indicate sample means for each layer.}
\label{fig:3}
\end{figure}

\begin{table}[!htbp]
\centering
\caption{AWPRI Scores, Layer Decomposition, and OLS Trend by Country, 2022}
\label{tab:2}
\small
\begin{tabular*}{\linewidth}{@{\extracolsep{\fill}}lllllll@{}}
\toprule
\textbf{Country} & \textbf{AWPRI} & \textbf{L1} & \textbf{L2} & \textbf{L3} & \textbf{Slope (per year)} & \textbf{p} \\
\midrule
China & 0.7027 & 0.5108 & 0.9995 & 0.5979 & +0.0026 & 0.054 ns \\
Vietnam & 0.6631 & 0.4810 & 0.8519 & 0.6565 & +0.0091 & 0.038 * \\
Thailand & 0.6116 & 0.5479 & 0.7291 & 0.5578 & +0.0033 & 0.304 ns \\
India & 0.5780 & 0.3238 & 0.5293 & 0.8810 & +0.0093 & < 0.001 *** \\
South Korea & 0.5453 & 0.6869 & 0.3171 & 0.6319 & +0.0102 & < 0.001 *** \\
United States & 0.5045 & 0.7570 & 0.3022 & 0.4544 & +0.0037 & 0.026 * \\
Brazil & 0.4919 & 0.6532 & 0.4379 & 0.3845 & $-$0.0008 & 0.616 ns \\
Canada & 0.4805 & 0.6167 & 0.2680 & 0.5567 & +0.0015 & 0.430 ns \\
Australia & 0.4788 & 0.7353 & 0.2205 & 0.4805 & +0.0013 & 0.353 ns \\
Mexico & 0.4741 & 0.5154 & 0.4960 & 0.4110 & +0.0003 & 0.807 ns \\
Spain & 0.4629 & 0.6797 & 0.1714 & 0.5376 & $-$0.0011 & 0.518 ns \\
Poland & 0.4520 & 0.6681 & 0.2797 & 0.4082 & +0.0065 & 0.001 ** \\
Nigeria & 0.4376 & 0.2327 & 0.5424 & 0.5376 & +0.0012 & 0.686 ns \\
Italy & 0.4354 & 0.5405 & 0.1899 & 0.5758 & +0.0068 & 0.005 ** \\
France & 0.4351 & 0.6077 & 0.2374 & 0.4603 & +0.0038 & 0.025 * \\
Argentina & 0.4270 & 0.5704 & 0.4169 & 0.2937 & $-$0.0036 & 0.209 ns \\
South Africa & 0.4204 & 0.4033 & 0.3705 & 0.4873 & $-$0.0002 & 0.935 ns \\
Netherlands & 0.3932 & 0.6850 & 0.0835 & 0.4111 & +0.0035 & 0.048 * \\
United Kingdom & 0.3843 & 0.5595 & 0.2253 & 0.3682 & +0.0043 & < 0.001 *** \\
New Zealand & 0.3784 & 0.6773 & 0.2424 & 0.2155 & +0.0004 & 0.882 ns \\
Denmark & 0.3640 & 0.6506 & 0.1182 & 0.3232 & +0.0046 & 0.023 * \\
Germany & 0.3617 & 0.4841 & 0.2008 & 0.4003 & $-$0.0004 & 0.638 ns \\
Kenya & 0.3386 & 0.1575 & 0.4515 & 0.4069 & $-$0.0010 & 0.729 ns \\
Sweden & 0.3161 & 0.5634 & 0.0166 & 0.3682 & +0.0033 & 0.064 ns \\
\bottomrule
\end{tabular*}
\par\smallskip
{\footnotesize\raggedright \textit{Note. Slope column reports the OLS trend coefficient (AWPRI $\sim$ year, 2010--2022) in index points per year, with its p value. A positive slope is movement towards higher exposure or weaker governance capacity relative to the panel. * p < 0.05; ** p < 0.01; *** p < 0.001; ns = non-significant.}\par}
\end{table}

\subsection{Cluster Structure}

Table 3 presents the partition k-means returns on the 2022 layer scores. The silhouette coefficient selects k = 2 (0.504), above k = 3 (0.444), k = 4 (0.330), k = 5 (0.326), k = 6 (0.341), and k = 7 (0.335). The Calinski--Harabasz index also peaks at k = 2 (25.09); the Davies--Bouldin index is marginally lower at k = 3 (0.849 against 0.851), which is the one diagnostic of the three that does not select the two-group solution. Our data support two groups and not the four tiers our initial analysis reported. The higher group (n = 6) comprises China, India, Kenya, Nigeria, Thailand, and Vietnam, with a mean composite of 0.555. The lower group (n = 18) holds the remaining countries with a mean composite of 0.434. Such a dichotomous separation is carried by L2 (higher group mean 0.684, lower group mean 0.255) and not by L1 (0.376 against 0.614). What k-means finds in this panel is a governance capacity split, and it is closer to an income split than to any welfare ordering.

\begin{table}[!htbp]
\centering
\caption{K-Means Partition of the 2022 Layer Scores (k = 2)}
\label{tab:3}
\small
\begin{tabular}{@{}>{\raggedright\arraybackslash}p{\dimexpr 0.1063\linewidth-2\tabcolsep\relax}>{\raggedright\arraybackslash}p{\dimexpr 0.5005\linewidth-2\tabcolsep\relax}>{\raggedright\arraybackslash}p{\dimexpr 0.0607\linewidth-2\tabcolsep\relax}>{\raggedright\arraybackslash}p{\dimexpr 0.1365\linewidth-2\tabcolsep\relax}>{\raggedright\arraybackslash}p{\dimexpr 0.1960\linewidth-2\tabcolsep\relax}@{}}
\toprule
\textbf{Group} & \textbf{Countries} & \textbf{n} & \textbf{Mean AWPRI} & \textbf{Mean L1 / L2 / L3} \\
\midrule
Higher-scoring group & China, India, Kenya, Nigeria, Thailand, Vietnam & 6 & 0.555 & 0.376 / 0.684 / 0.606 \\
Lower-scoring group & Argentina, Australia, Brazil, Canada, Denmark, France, Germany, Italy, Mexico, Netherlands, New Zealand, Poland, South Africa, South Korea, Spain, Sweden, United Kingdom, United States & 18 & 0.434 & 0.614 / 0.255 / 0.432 \\
\bottomrule
\end{tabular}
\par\smallskip
{\footnotesize\raggedright \textit{Note. Groups are the k-means assignment on L1, L2, and L3 in 2022, with k selected on the silhouette coefficient. Group names state only which group holds the higher mean composite score and carry no tier interpretation.}\par}
\end{table}

Figure 4 presents the cluster diagnostics. The elbow method applied to within-cluster sum of squares does not identify a clear inflection, which is what a 24-point cross-section in three dimensions will usually suggest, and we therefore select on the silhouette coefficient alone. The silhouette is highest at k = 2 (0.504) and drops thereafter, and the initial four-tier typology is withdrawn. We note that a two-group solution on 24 countries is a weak structure and that the higher group holds the five countries with the weakest governance capacity and Kenya, the seventh weakest.

K-means partitions the panel into two groups using all three layer scores. Comparing those same two groups on the composite score, however, neither the Kruskal--Wallis test (H = 3.738, p = 0.053) nor the Mann--Whitney U test (U = 83, p = 0.056) reaches the conventional level of statistical significance. Such an outcome happens because the composite averages a layer on which the groups differ notably (L2) with a layer on which they differ just as substantially but in the opposite direction (L1). We find that the group with the higher L2 score has the lower L1 score, and \textit{vice versa}, so the two differences partly cancel out once averaged into one composite number. Our empirical findings fail to validate such a partition. Differences across the six World Bank regions in 2022 reach statistical significance at the 5\% level on the composite (H = 12.680, p = 0.027), driven by L2 (H = 15.116, p = 0.010) and not by L1 (H = 10.839, p = 0.055) or L3 (H = 8.828, p = 0.116).

\begin{figure}[!htbp]
\centering
\includegraphics[width=0.955\linewidth]{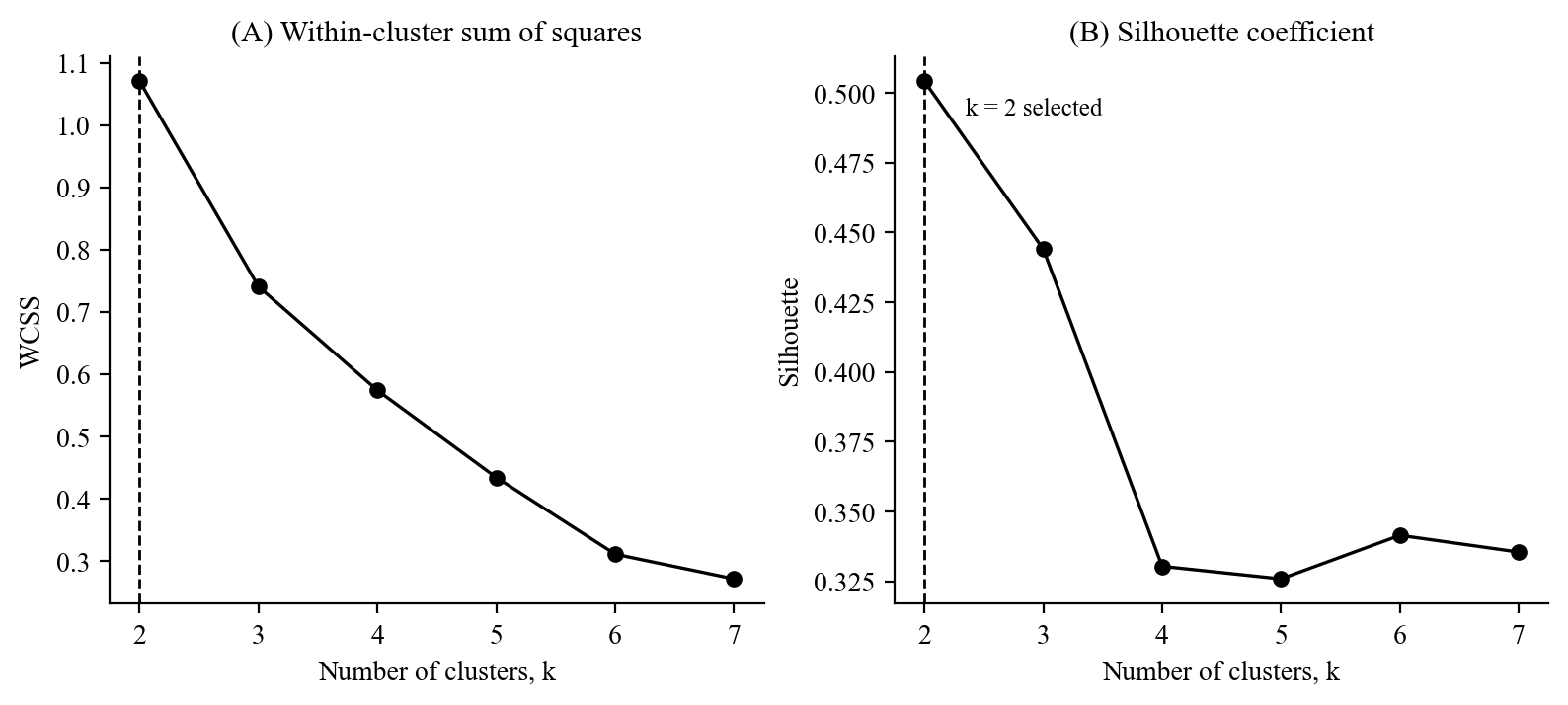}
\caption{Cluster Diagnostics. (A) Elbow plot of within-cluster sum of squares by k. (B) Silhouette coefficient by k. Dashed vertical line at k = 2 indicates the selected solution.}
\label{fig:4}
\end{figure}

\subsection{Temporal Dynamics, 2010--2022}

Figure 5 presents the AWPRI trajectory of every country in the panel, one panel to a country and ordered by the size of the trend slope. The Slope column in Table 2 reports OLS trend coefficients and p values for all 24 countries. A total of 10 of the 24 display a slope distinguishable from zero over 2010--2022 (p < 0.05), and all 10 are positive. No country records a statistically significant negative slope. Among them, South Korea records the steepest slope ($\beta$ = 0.0102 per year, p < 0.001), followed by India ($\beta$ = 0.0093, p < 0.001) and Vietnam ($\beta$ = 0.0091, p = 0.038). A total of 14 countries record slopes that are not distinguishable from zero, including six with negative point estimates (Argentina, Brazil, Germany, Kenya, South Africa, and Spain). The panel mean AWPRI grows from 0.430 in 2010 to 0.464 in 2022. Because each variable is normalised within each year, the panel mean measures where the 24 countries sit between that year's lowest and highest values, and the increase does not show that the panel as a whole moved towards higher exposure or weaker governance capacity.

When each variable is instead normalised against bounds pooled across all 13 years, 21 of the 24 slopes are positive and distinguishable from zero (p < 0.05), and no slope is negative and distinguishable from zero. All 10 countries with a significant slope under within-year normalisation keep a significant slope, and 11 more join them, namely Australia, Brazil, Canada, China, Germany, Mexico, New Zealand, South Africa, Spain, Sweden, and Thailand. The exceptions are Argentina (p = 0.108), Kenya (p = 0.533), and Nigeria (p = 0.069). The sign of the slope agrees under both normalisations for 18 of the 24 countries, and the Spearman correlation between the two sets of slopes is 0.879. Under pooled bounds the panel mean AWPRI grows from 0.397 in 2010 to 0.479 in 2022, against 0.430 to 0.464 under within-year normalisation, and most of that increase comes from L3, which grows from 0.332 to 0.551, against 0.498 to 0.516 for L1 and 0.362 to 0.371 for L2. The L3 slope is distinguishable from zero for 23 of the 24 countries, whereas the mean of L1 and L2 increases significantly for nine countries and decreases significantly for four.

\begin{figure}[!htbp]
\centering
\includegraphics[width=0.784\linewidth]{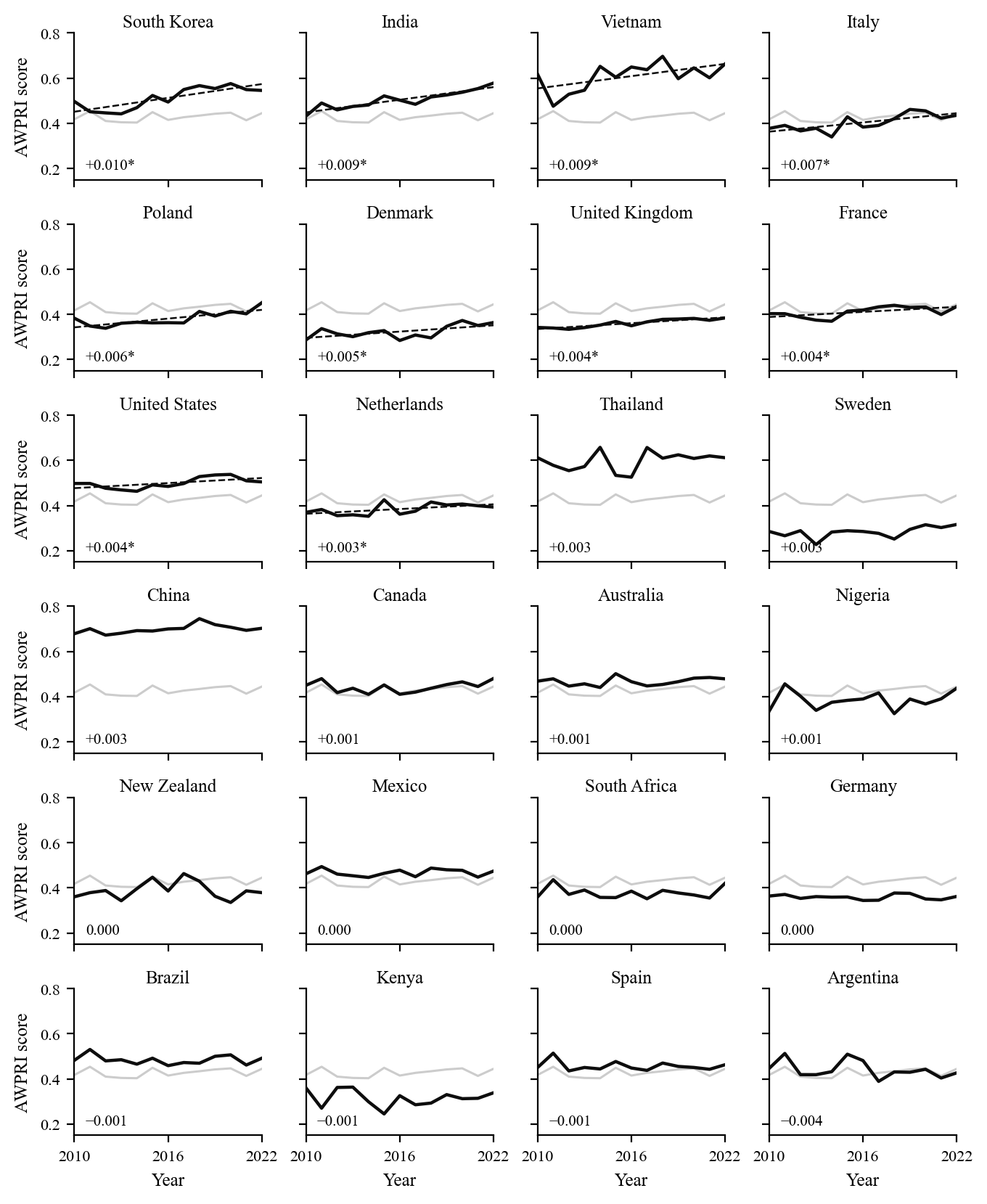}
\caption{AWPRI Temporal Trajectories, 2010--2022. One panel per country, ordered by the size of the OLS trend slope. The heavy line is the country's AWPRI score and the light line repeated in every panel is the 24-country median for the same year. The dashed line is the fitted OLS trend and is drawn only where the slope is distinguishable from zero at p < 0.05. The number in the lower left of each panel is the slope in AWPRI points per year, with an asterisk where p < 0.05. All panels share the same axes.}
\label{fig:5}
\end{figure}

Figure 6 presents OLS trend slope coefficients for all 24 countries with 95\% confidence intervals. The countries with significant positive slopes are not concentrated in either k-means group. Two of the 10 sit in the higher group (India and Vietnam) and eight sit in the lower group, including the United Kingdom ($\beta$ = 0.0043, p < 0.001) and Denmark ($\beta$ = 0.0046, p = 0.023).

\begin{figure}[!htbp]
\centering
\includegraphics[width=0.911\linewidth]{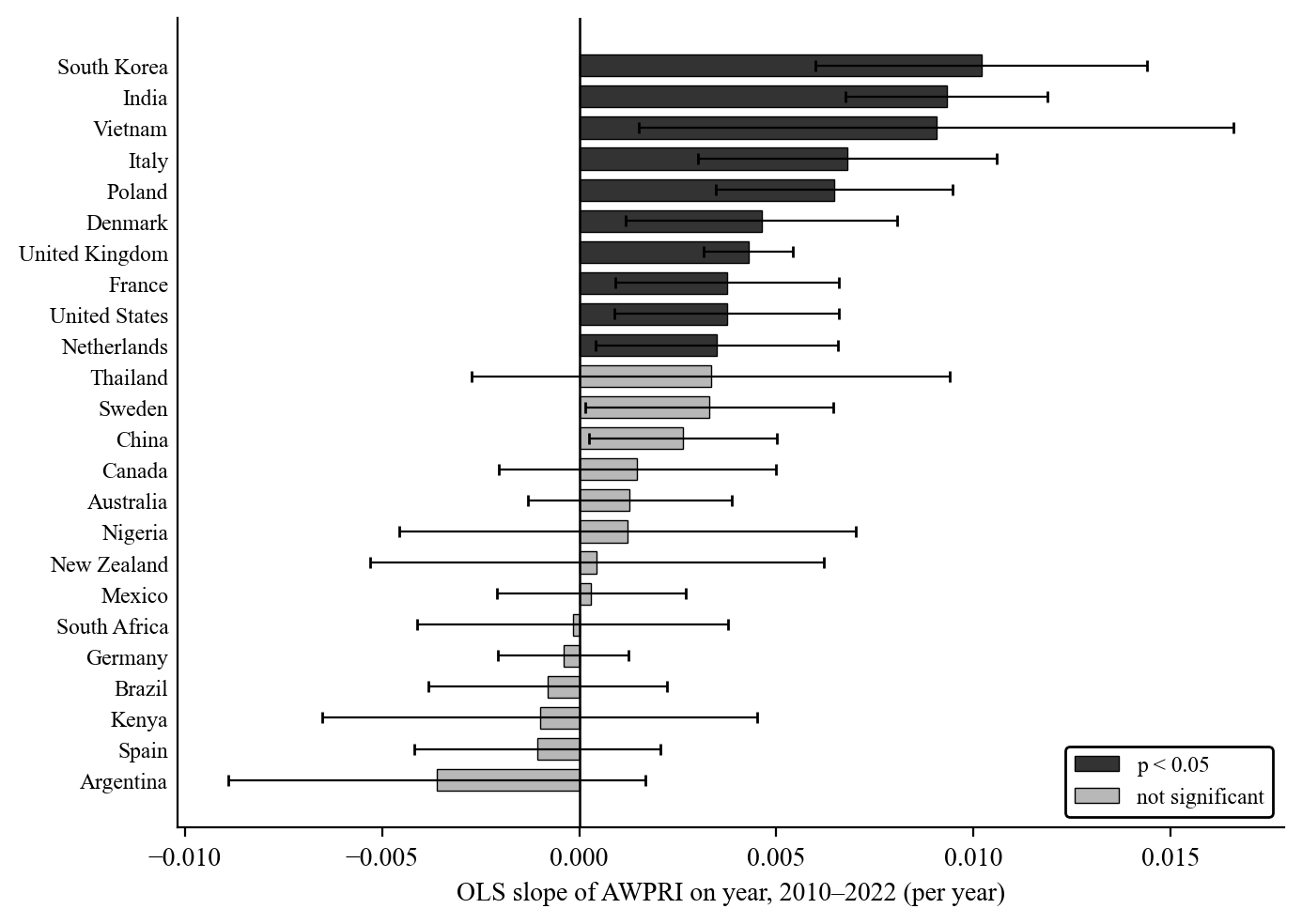}
\caption{OLS Trend Slope Coefficients by Country, 2010--2022. Bars indicate $\beta$ coefficients from country-level OLS regressions of AWPRI on year. Error bars indicate 95\% confidence intervals. Countries ordered by slope. Dark bars indicate a slope distinguishable from zero at p < 0.05.}
\label{fig:6}
\end{figure}

\subsection{External Validation and Event Study}

Table 4 reports the external validation against the WAP's Animal Protection Index 2020 (n = 24), on the overall grade and on the farm animal grade. The composite correlates with neither the overall grade ($\rho$ = $-$0.264, p = 0.213) nor the farm animal grade ($\rho$ = $-$0.301, p = 0.153), and the composite correlates with neither grade net of log gross domestic product per capita (partial $\rho$ = $-$0.222, p = 0.309, and partial $\rho$ = $-$0.269, p = 0.214). Animal-use exposure correlates with the overall grade (L1 $\rho$ = $-$0.485, p = 0.016; partial $\rho$ = $-$0.461, p = 0.027). Against the farm animal grade, the L1 correlation has the same sign but is smaller and not statistically significant ($\rho$ = $-$0.392, p = 0.059; partial $\rho$ = $-$0.355, p = 0.096), so the L1 result holds for the overall grade only. Governance capacity ($\rho$ = $-$0.106 and $-$0.136) and AI amplification ($\rho$ = $-$0.208 and $-$0.302) correlate with neither grade (all p $\geq$ 0.151). Among the 12 constituent variables, rule of law correlates most strongly with both grades ($\rho$ = $-$0.684 and $-$0.780, both p < 0.001) and remains statistically significant after controlling for income (partial $\rho$ = $-$0.530, p = 0.009, and partial $\rho$ = $-$0.627, p = 0.001). Civil liberties also correlates with both grades ($\rho$ = $-$0.594, p = 0.002, and $\rho$ = $-$0.753, p < 0.001), but after controlling for income only the correlation with the farm animal grade remains statistically significant (partial $\rho$ = $-$0.349, p = 0.102, and partial $\rho$ = $-$0.565, p = 0.005). Two variables correlate positively with a WAP grade, the animal share of protein supply with both grades ($\rho$ = +0.452, p = 0.027, and $\rho$ = +0.549, p = 0.005) and farmed animals slaughtered per capita with the farm animal grade ($\rho$ = +0.494, p = 0.014). Wealthier countries tend to consume more animal protein and enact more welfare legislation simultaneously; controlling for income, the three positive correlations drop to between +0.053 and +0.256 (all p $\geq$ 0.239). The two instruments therefore appear to measure different constructs. The WAP grade evaluates a country's existing legislation, whereas the AWPRI weighs that legislative record against the scale of animal use in the country's food system; as a result, a country may receive a favourable WAP grade while still recording a high exposure score.

Figure 7 and Table 5 present the event study around \textit{Regulation (EU) 2017/625}, which became legally binding on member states from 14 December 2019. The treated group is the eight EU member states in the panel plus the United Kingdom (n = 9); the control group is the remaining 15 countries. The outcome variable is the composite score excluding the two FAOLEX-derived legislation variables. This exclusion is necessary because an EU regulation, once enacted, is itself recorded in FAOLEX; retaining these variables would therefore inflate the outcome independent of any substantive change in enforcement. Event time runs from k = $-$9 (2010) to k = +3 (2022) with k = $-$1 (2018) omitted as the reference, and standard errors are clustered by country.

None of the eight pre-treatment coefficients (leads) differs significantly from zero (p = 0.408--0.863; point estimates $-$0.004 to 0.015), providing no evidence against the parallel-trends assumption.~None of the four post-treatment coefficients differs significantly from zero either: at k = +3 (2022), the estimate is 0.012 (SE = 0.017, p = 0.497, 95\% CI [$-$0.022, 0.045]). This null result holds for each subcomponent of the composite, namely the governance layer excluding the two legislation variables (k = +3: 0.010, SE = 0.015, p = 0.497), the exposure layer (k = +3: $-$0.008, SE = 0.007, p = 0.260), and the AI amplification layer (k = +3: 0.033, SE = 0.052, p = 0.530). The estimates are unchanged to three decimal places when the United Kingdom is dropped from the sample, and the null result remains consistent when the event is dated to 2020 instead of 2019 (k = +2: 0.005, SE = 0.012, p = 0.694).

The legislation-purged composite shows no detectable change when the nine treated countries become subject to a binding official-controls regulation; we report this outcome as a null result. Three interpretations are available, and the present design cannot distinguish among them. First, the regulation may not have changed the conditions that the 10 variables of the legislation-purged composite measure within the four-year post-treatment window. Second, the legislation-purged composite may be insensitive to the type of change the regulation leads to. Third, an effect may exist but be too small for a 24-country panel to detect. Three of the four post-treatment years align with the outbreak of the COVID-19 pandemic, which affected food systems, research output, and enforcement capacity unevenly across Europe and the rest of the sample; as a result, the post-treatment estimates address the combined effect of the regulation and the pandemic, and our research design fails to separate the two. With only 24 clusters, cluster-robust standard errors are anti-conservative, and EU membership correlates with income, governance capacity, and most of the variables composing the index.

\begin{figure}[!htbp]
\centering
\includegraphics[width=0.852\linewidth]{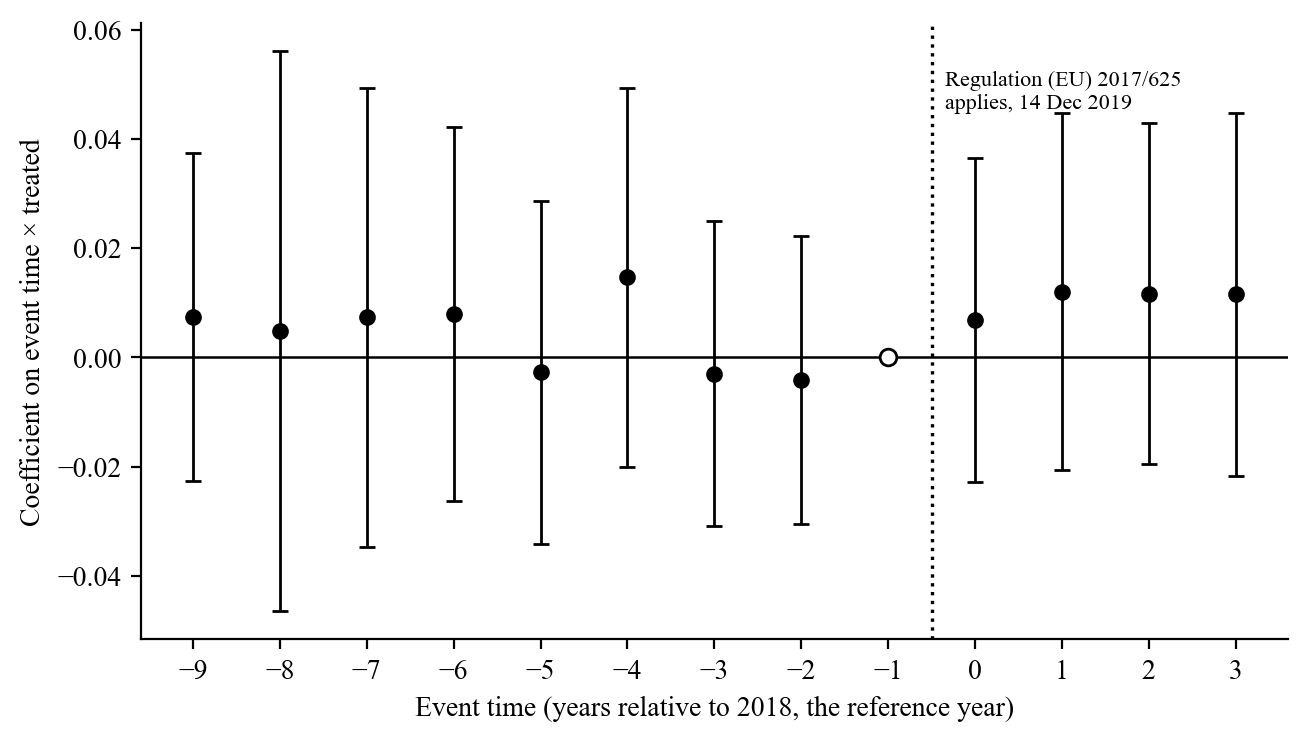}
\caption{Event Study around Regulation (EU) 2017/625. Points indicate event-time coefficients on the legislation-purged composite; bars indicate 95\% confidence intervals. The open marker at k = $-$1 (2018) is the omitted reference year. Dotted vertical line marks the application date, 14 December 2019.}
\label{fig:7}
\end{figure}

\begin{table}[!htbp]
\centering
\caption{External Validation of the 2020 AWPRI against the WAP's Animal Protection Index 2020, Overall Grade and Farm Animal Grade (n = 24)}
\label{tab:4}
\footnotesize
\setlength{\tabcolsep}{3pt}
\begin{tabular}{@{}>{\raggedright\arraybackslash}p{\dimexpr 0.1600\linewidth-2\tabcolsep\relax}>{\centering\arraybackslash}p{\dimexpr 0.1250\linewidth-2\tabcolsep\relax}>{\centering\arraybackslash}p{\dimexpr 0.0850\linewidth-2\tabcolsep\relax}>{\centering\arraybackslash}p{\dimexpr 0.1100\linewidth-2\tabcolsep\relax}>{\centering\arraybackslash}p{\dimexpr 0.1000\linewidth-2\tabcolsep\relax}>{\centering\arraybackslash}p{\dimexpr 0.1250\linewidth-2\tabcolsep\relax}>{\centering\arraybackslash}p{\dimexpr 0.0850\linewidth-2\tabcolsep\relax}>{\centering\arraybackslash}p{\dimexpr 0.1100\linewidth-2\tabcolsep\relax}>{\centering\arraybackslash}p{\dimexpr 0.1000\linewidth-2\tabcolsep\relax}@{}}
\toprule
\textbf{AWPRI quantity} & \textbf{Spearman $\rho$, overall} & \textbf{p} & \textbf{\textit{Partial $\rho$, overall}} & \textbf{\textit{p (partial)}} & \textbf{Spearman $\rho$, farm} & \textbf{p} & \textbf{\textit{Partial $\rho$, farm}} & \textbf{\textit{p (partial)}} \\
\midrule
AWPRI composite & $-$0.2640 & 0.213 & $-$0.2217 & 0.309 & $-$0.3012 & 0.153 & $-$0.2691 & 0.214 \\
L1, Animal-use exposure & $-$0.4845 & 0.016 & $-$0.4610 & 0.027 & $-$0.3915 & 0.059 & $-$0.3552 & 0.096 \\
L2, Governance capacity & $-$0.1055 & 0.624 & $-$0.0335 & 0.879 & $-$0.1357 & 0.527 & $-$0.0583 & 0.792 \\
L3, AI amplification & $-$0.2078 & 0.330 & $-$0.1915 & 0.381 & $-$0.3021 & 0.151 & $-$0.3149 & 0.143 \\
Rule of law (\textit{VAR\_04}) & $-$0.6839 & < 0.001 & $-$0.5301 & 0.009 & $-$0.7803 & < 0.001 & $-$0.6268 & 0.001 \\
Civil liberties (\textit{VAR\_10}) & $-$0.5940 & 0.002 & $-$0.3493 & 0.102 & $-$0.7533 & < 0.001 & $-$0.5652 & 0.005 \\
Animal share of protein supply (\textit{VAR\_07}) & +0.4521 & 0.027 & +0.0529 & 0.810 & +0.5493 & 0.005 & +0.1171 & 0.595 \\
Farmed animals slaughtered per capita (\textit{VAR\_01}) & +0.2069 & 0.332 & $-$0.0949 & 0.667 & +0.4935 & 0.014 & +0.2559 & 0.239 \\
\bottomrule
\end{tabular}
\par\smallskip
{\footnotesize\raggedright \textit{Note. Spearman rank correlations between 2020 AWPRI quantities and the WAP's Animal Protection Index 2020 grades (WAP, 2020). The farm animal grade is the grade on the indicator for protecting animals used in farming. WAP grades are scored so that a higher number is a better grade, so a negative correlation is the direction a working index should produce. Partial correlations are net of log gross domestic product per capita. The composite returns a null against both grades; L1 correlates with the overall grade and not, at p < 0.05, with the farm animal grade.}\par}
\end{table}

\begin{table}[!htbp]
\centering
\caption{Event-Study Coefficients around \textit{Regulation (EU) 2017/625}, Legislation-Purged Outcomes (N = 312)}
\label{tab:5}
\small
\begin{tabular*}{\linewidth}{@{\extracolsep{\fill}}llllll@{}}
\toprule
\textbf{Event time (year)} & \textbf{\textit{$\beta$}} & S.E. & \textbf{\textit{p}} & \textbf{\textit{95\% CI}} & \textbf{Period} \\
\midrule
k = $-$9 (2010) & +0.0074 & 0.0153 & 0.631 & [$-$0.0227, +0.0374] & Pre \\
k = $-$8 (2011) & +0.0048 & 0.0262 & 0.856 & [$-$0.0465, +0.0560] & Pre \\
k = $-$7 (2012) & +0.0073 & 0.0214 & 0.732 & [$-$0.0346, +0.0493] & Pre \\
k = $-$6 (2013) & +0.0080 & 0.0174 & 0.648 & [$-$0.0262, +0.0422] & Pre \\
k = $-$5 (2014) & $-$0.0028 & 0.0160 & 0.863 & [$-$0.0342, +0.0287] & Pre \\
k = $-$4 (2015) & +0.0146 & 0.0177 & 0.408 & [$-$0.0200, +0.0493] & Pre \\
k = $-$3 (2016) & $-$0.0030 & 0.0142 & 0.835 & [$-$0.0308, +0.0249] & Pre \\
k = $-$2 (2017) & $-$0.0041 & 0.0135 & 0.759 & [$-$0.0305, +0.0222] & Pre \\
k = 0 (2019) & +0.0068 & 0.0151 & 0.651 & [$-$0.0228, +0.0365] & Post \\
k = +1 (2020) & +0.0120 & 0.0166 & 0.469 & [$-$0.0206, +0.0447] & Post \\
k = +2 (2021) & +0.0117 & 0.0159 & 0.464 & [$-$0.0195, +0.0428] & Post \\
k = +3 (2022) & +0.0115 & 0.0170 & 0.497 & [$-$0.0218, +0.0448] & Post \\
\bottomrule
\end{tabular*}
\par\smallskip
{\footnotesize\raggedright \textit{Note. Two-way fixed effects event study with country and year effects, k = $-$1 (2018) omitted as the reference, standard errors clustered by country. Outcomes are purged of the two FAOLEX legislation variables, which an EU regulation would move mechanically. No coefficient in any specification reaches p < 0.05. The initial difference-in-differences estimate is withdrawn for the reasons stated in Section 3.6.}\par}
\end{table}

\subsection{PCA}

Figure 8 shows the PCA of the standardised 12-variable 2022 cross-section. Figure 8(A) presents the scree plot. PC1 accounts for 47.8\% of total variance and PC2 for 19.8\%; five components are required to reach 90.8\% cumulative variance, as indicated by the dotted horizontal line. The findings show that the 12 indicators are not reducible to a single dimension. Figure 8(B) presents the PC1--PC2 biplot. PC1 loads positively on the five governance capacity variables (\textit{civil\_\allowbreak liberties\_\allowbreak risk} 0.374, \textit{rule\_\allowbreak of\_\allowbreak law\_\allowbreak risk} 0.367, \textit{civic\_\allowbreak space\_\allowbreak risk} 0.314, \textit{welfare\_\allowbreak legislation\_\allowbreak stock\_\allowbreak risk} 0.288, \textit{welfare\_\allowbreak legislation\_\allowbreak flow\_\allowbreak risk} 0.265) and negatively on three of the four exposure variables (\textit{animal\_\allowbreak protein\_\allowbreak share} $-$0.362, \textit{meat\_\allowbreak supply\_\allowbreak kg} $-$0.308, \textit{farmed\_\allowbreak animals\_\allowbreak per\_\allowbreak capita} $-$0.262), which means the first component is the opposition between animal-use exposure and governance capacity and not a general risk factor. The six countries in the higher k-means group occupy the positive region of PC1, and 17 of the remaining 18 countries occupy the negative region. South Korea ($-$0.01), South Africa ($-$0.11), and Mexico (0.27) have PC1 scores close to zero and therefore do not project strongly into either region. Our three-layer composition of the AWPRI is therefore not redundant with a single principal component, and the composite is averaging two quantities.

\begin{figure}[!htbp]
\centering
\includegraphics[width=1.000\linewidth]{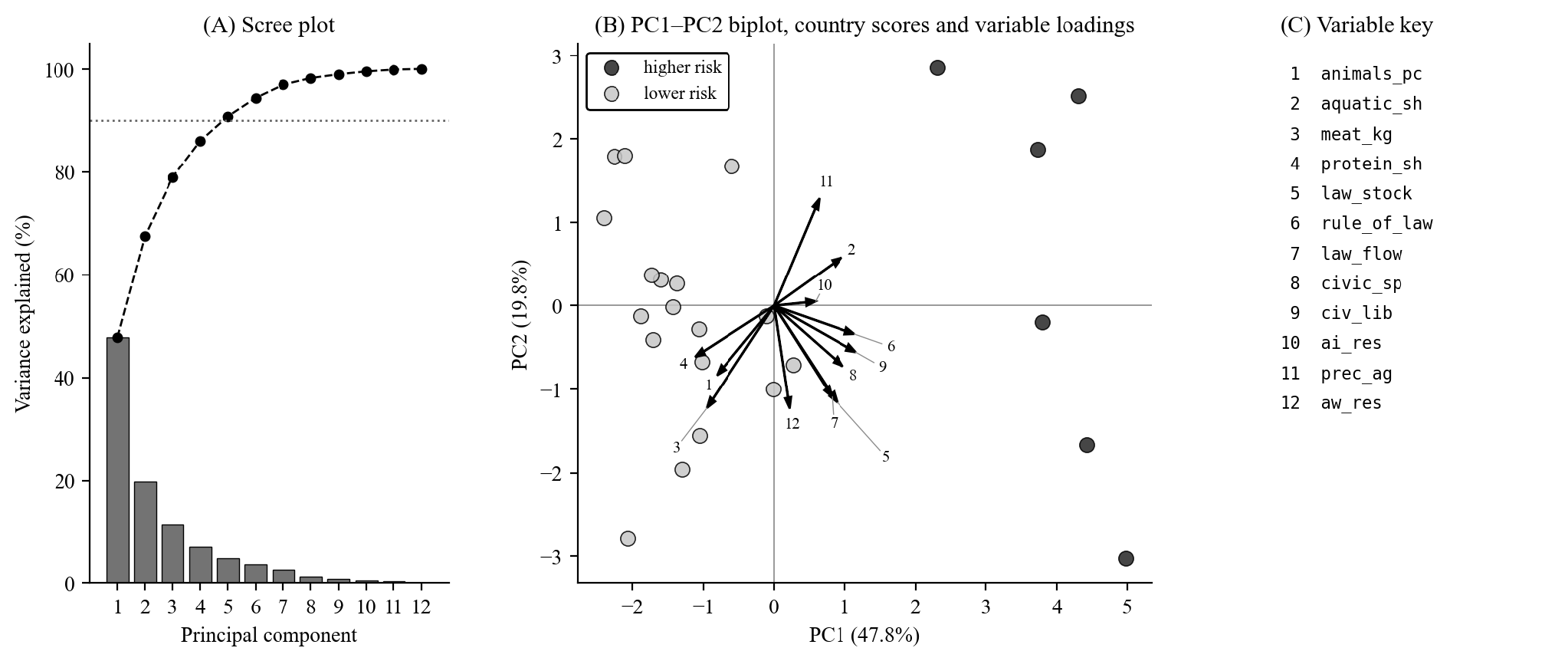}
\caption{PCA, 12-Variable Cross-Section, 2022. (A) Scree plot. Bars indicate variance explained by component, the dashed line indicates cumulative variance, and the dotted line marks 90\%. (B) PC1--PC2 biplot. Arrows indicate variable loadings; country markers shaded by k-means group.}
\label{fig:8}
\end{figure}

\subsection{Robustness of the Construction}

Table A3 (Appendix) presents the 46 rebuilds described in Section 3.7. Changing the normalisation affects the global ranking least. The rank correlation with the index we built is 0.994 when each variable is z-score normalised within each year (no country changes rank by more than two positions), 0.982 when each variable is min-max normalised over all 13 years (three positions), and 0.985 when each variable is z-score normalised over all 13 years (four positions). Reweighting the layers changes the ranking more than changing the normalisation. The rank correlation is 0.914 when the exposure layer's weight is doubled (Nigeria changes rank by nine positions) and 0.926 when the governance layer's weight is doubled (Argentina, Kenya, and Nigeria by five positions each). When one variable is removed, the rank correlation is between 0.865 and 0.990 and is lowest for the animal welfare research base (0.865, the United States by 10 positions). Removing a layer (L1, L2, or L3) changes the ranking more, because each layer is one third of the index, and removing the governance layer changes the ranking most (0.681, Nigeria by 10 positions and Spain by nine). Removing one country tests whether the minimum and maximum of each variable depend on which countries are in the panel, and the other 23 countries barely change rank. In 18 of the 24 rankings rebuilt without one country, the rank correlation is exactly 1.000 and no country changes rank, and the lowest rank correlation is 0.994, when China is removed (Mexico by two positions). We report the rank correlation and the largest change in rank for each rebuilt ranking and set no threshold above which the index should be considered robust.

There is no data imputation to check. Every one of the 3,744 values in the panel is observed in a named public source for the country and year to which the value is attached. A country-year with any variable missing from its public source is excluded from the panel instead of being filled with an estimated value, and our Python-based assembly script enforces this rule by halting with an error if a missing value is ever encountered.

\subsection{Trend Projections to 2030}

Our projections are OLS linear trend extrapolations of each country's 2010--2022 series, for the reasons given in Section 3.5. This method assumes the fitted slope continues to 2030, an assumption our data support unevenly. Here, 10 of the 24 slopes are distinguishable from zero at p < 0.05, while the other 14 projections extrapolate a slope the data cannot separate from a flat line. Table S1 (Supplementary Material) reports the slope p value alongside every projection for this reason. Because each variable is normalised within each year, a 2030 projection extends a country's 2010--2022 position relative to the other 23 countries and is not a projected level of animal-use exposure, governance capacity, or research composition. The 95\% prediction interval covers the uncertainty of the fitted trend and the scatter of the observed scores around the fitted trend, and does not cover changes in which countries set the minimum and maximum of each variable in future years.

Figure 9 presents the projections to 2030. Table S1 (Supplementary Material) reports point forecasts and 95\% prediction intervals for all 24 countries. Of the 10 countries whose slope is distinguishable from zero, all 10 are projected towards upward trajectories to higher scores, and the largest projected 2030 values are China (0.736, slope p = 0.054), Vietnam (0.735, p = 0.038), South Korea (0.655, p < 0.001), Thailand (0.644, p = 0.304), and India (0.635, p < 0.001). Of these five, only Vietnam, South Korea, and India rely on a slope distinguishable from zero. At the lower end, Kenya (0.301, p = 0.729), Sweden (0.329, p = 0.064), and Germany (0.353, p = 0.638) hold the lowest projected values by 2030, and none of the three relies on a statistically significant slope either. The six countries with negative point slopes (Argentina, Brazil, Germany, Kenya, South Africa, and Spain) are projected to record lower scores in 2030 than in 2022, and none of the six slopes reaches statistical significance. Canada, Mexico, and Nigeria are also projected below their 2022 scores despite positive point slopes (Table S1), because a projection extends the fitted line and not the observed 2022 score. We therefore make no claim about which countries will record higher or lower scores by 2030.

\begin{figure}[!htbp]
\centering
\includegraphics[width=0.981\linewidth]{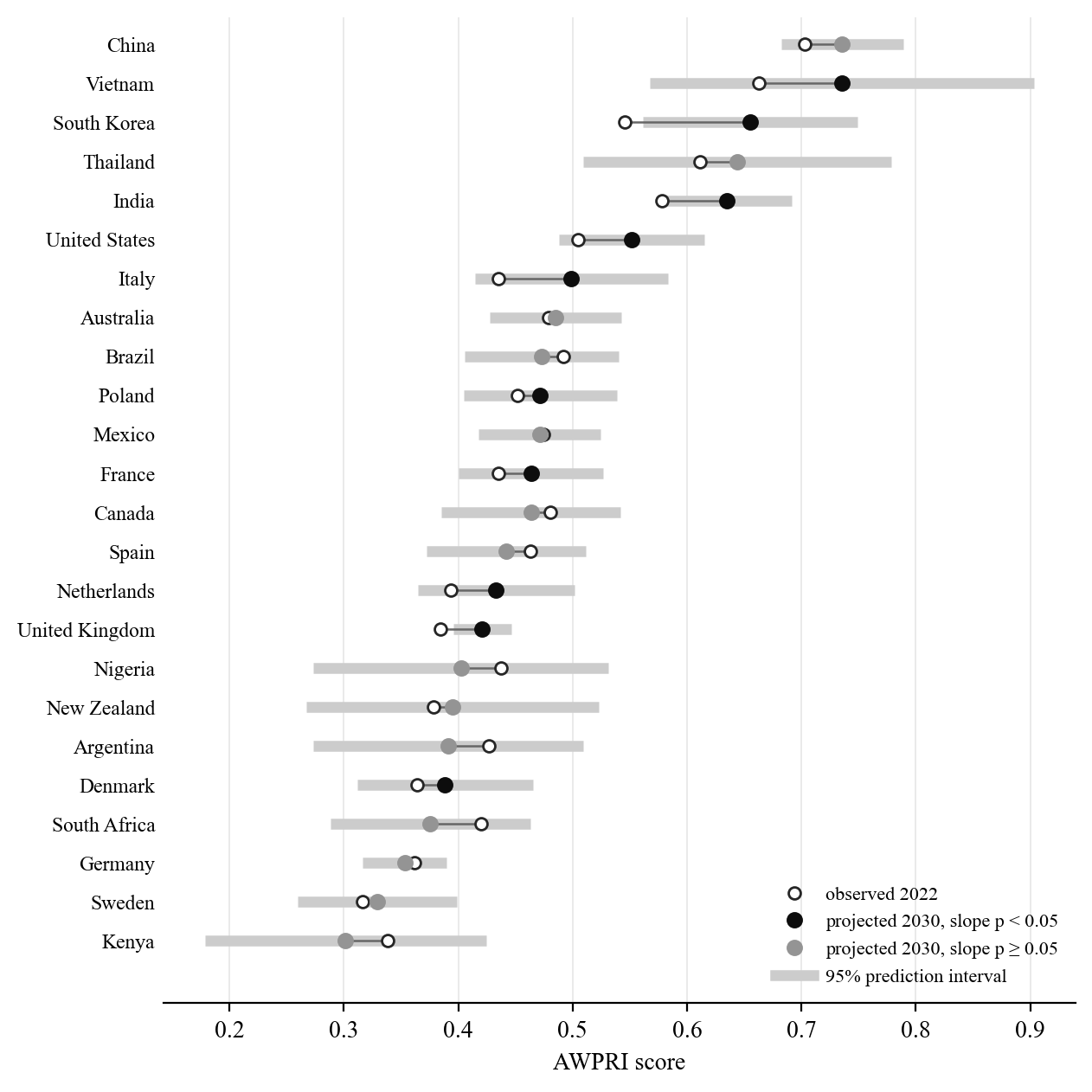}
\caption{AWPRI Trend Projections to 2030, All 24 Countries. Open markers indicate the observed 2022 score and filled markers the 2030 projection, joined by a line. Filled black markers indicate countries whose trend slope is distinguishable from zero at p < 0.05 and filled grey markers those whose slope is not. Grey bars indicate the 95\% prediction interval around the 2030 projection. Countries are ordered by projected 2030 score.}
\label{fig:9}
\end{figure}

\section{Discussion}

\subsection{Exposure and Governance Capacity Are Separate Quantities}

Our initial analysis reported AI amplification as the dominant contributor to composite risk. On the rebuilt panel our empirical findings show otherwise. L1 holds the highest full-panel mean (0.532) and L3 the middle one (0.436), and in the 2022 cross-section the L3 and L1 difference does not reach statistical significance (W = 86, p = 0.069).

What our rebuilt panel does show is that animal-use exposure and governance capacity are separate quantities that the AWPRI composite averages altogether. PC1 loads positively on all five governance capacity variables and negatively on three of the four exposure variables, so the first principal component is the opposition between them. In 2022, L1 is the highest of the three layers for 16 of the 24 countries. The countries that sit low on the composite do so for two different reasons that the composite does not distinguish. Sweden (0.316) combines the strongest governance capacity in the panel (L2 = 0.017) with high exposure (L1 = 0.563). Kenya (0.339) sits low on exposure (L1 = 0.158) with weak governance capacity (L2 = 0.452). We report the layer scores alongside the composite throughout this research paper and in our interactive dashboard, so as to clearly present the change in which layer's score primarily contributes to a change in the composite average.

The event study around \textit{Regulation (EU) 2017/625} returns no effect on the legislation-purged composite or on any of the three layers at any event time. The null result concerns animal-use exposure, the three governance capacity variables that are not drawn from FAOLEX, and research composition, and the null result does not indicate whether the regulation changed inspection coverage, non-compliance, or the welfare of animals in the treated countries. Instead, the null result shows the limit of what a 24-country index built on annual public statistics can detect within the 2019--2022 post-treatment window, of which 2020, 2021, and 2022 were pandemic years.

\subsection{Regional Patterns and Country Cases}

Composite scores vary by world region in 2022, with East Asia and Pacific recording the highest regional mean (0.563) and Sub-Saharan Africa and Europe and Central Asia the lowest (0.399 and 0.401, respectively; Kruskal--Wallis H = 12.680, p = 0.027). This regional pattern is attributable almost entirely to governance capacity. L2 differs notably by region (H = 15.116, p = 0.010), while exposure (L1, p = 0.055) and AI amplification (L3, p = 0.116) do not. Because the composite closely tracks the L2 ranking of countries, the regional difference on the composite is a testament to the regional difference on L2. We do not believe this observation represents the regional nuances. This is because the six regions we used for empirical analysis hold between one and nine countries each, and South Asia is represented by India alone. Therefore, our empirical findings are shaped by the limitation of the panel dataset (such as by limited coverage of countries) that we have and do not necessarily present geographical/regional representation.

Relevant scholarship widely acknowledges the United Kingdom as a global leader in animal welfare legislation (H\aa{}rstad, 2024; WAP, 2020). In the rebuilt panel the United Kingdom records a composite of 0.384, which places it 19 of 24 and not 24 of 24, the lowest score in the panel, and its slope over 2010--2022 is positive and significant ($\beta$ = 0.0043, p < 0.001). We do not present this finding as the index performing as designed. The enforcement gap documented in Section 2, in which fewer than 2.5\% of farms in England were inspected in 2024 and fewer than 1\% of recorded violations resulted in prosecution (Animal Law Foundation, 2024), is not visible anywhere in the index, because the index contains no inspection or prosecution series. A country can hold the strongest welfare statutes in the sample and a failing inspectorate, and the AWPRI would not separate the two. That is a limitation of the instrument as stated in Section 6.

In North America, the United States records the highest L1 score in the panel (0.757) and an L2 score of 0.302, below the panel mean (0.362), and the United States composite of 0.505 places the country 6 of 24, with a positive and significant slope ($\beta$ = 0.0037, p = 0.026). Canada records a composite of 0.481 (L1 = 0.617, L2 = 0.268, L3 = 0.557) and a slope that is not distinguishable from zero ($\beta$ = 0.0015, p = 0.430). The United States L3 score of 0.454 does not come from AI research. The AI share of United States works is 2.82\%, against a panel median of 3.71\%, and United States precision agriculture research intensity is the lowest in the panel (1.68 works per 10,000, against a median of 3.41). The United States L3 score comes instead from the smallest animal welfare research base in the panel (3.75 works per 10,000, against a median of 7.87), which is why removing that variable moves the United States 10 places (Table A3, Appendix).

In Latin America, Brazil (0.492), Mexico (0.474), and Argentina (0.427) record no slope distinguishable from zero under within-year normalisation. The three countries record L1 scores between 0.515 and 0.653, L2 scores between 0.417 and 0.496, which indicate weaker governance capacity than every North American and European country in the panel, and L3 scores between 0.294 and 0.411, below the panel mean (0.475). Argentina records the lowest AI share in the panel (1.24\%) and moves six places when AI research intensity is removed (Table A3, Appendix). Under pooled bounds, the slopes of Brazil and Mexico become distinguishable from zero and the slope of Argentina does not (p = 0.108).

In Asia, the five highest composite scores in the panel all belong to Asian countries, namely China (0.703), Vietnam (0.663), Thailand (0.612), India (0.578), and South Korea (0.545), and the five countries reach those positions by three routes. China, Vietnam, and Thailand record the three weakest L2 scores in the panel (0.999, 0.852, and 0.729). India records the highest L3 score in the panel (0.881), with an L2 score of 0.529. South Korea records an L2 score of 0.317, close to those of the United States (0.302) and Canada (0.268), and reaches the top five on a high L1 score (0.687) and the third-highest L3 score (0.632). South Korea, India, and Vietnam also record the three steepest slopes in the panel (Section 4.5).

\subsection{Temporal Trajectories}

Our temporal trend analysis shows changes in one direction only. A total of 10 of the 24 countries record a slope distinguishable from zero over 2010--2022, and all 10 are positive. They are South Korea, India, Vietnam, Italy, Poland, Denmark, the United Kingdom, France, the United States, and the Netherlands, which is a list that includes countries from both k-means groups and from both high-income and lower-middle-income categories. Under within-year normalisation, a positive slope indicates that a country moved up relative to the other 23 countries, and the increase in the panel mean from 0.430 to 0.464 does not on its own show a change in the shape of the distribution within each year. Under pooled bounds, 21 of the 24 slopes are positive and distinguishable from zero and most of the increase in the panel mean comes from L3 (Section 4.5), so the direction of change holds when the bounds are fixed across years, and what changes most is the composition of national research effort.

\subsection{Interpreting the AI Amplification Layer}

L3 changes more within countries than the other two layers (mean within-country SD = 0.088, against 0.020 for L1 and 0.023 for L2), and L3 accounts for most of the increase in the panel mean under pooled bounds (Section 4.5). In 2022, the four countries with the largest AI shares of national works, namely India (8.74\%), Vietnam (7.37\%), South Korea (7.26\%), and China (6.76\%), record the four highest L3 scores. The animal welfare research base also shapes L3. The United States records an L3 score of 0.454 with an AI share below the panel median and the smallest welfare research base in the panel, whereas New Zealand records the lowest L3 score (0.215) with the largest welfare research base (24.60 works per 10,000). Kenya records a precision agriculture research intensity of 11.75 works per 10,000, more than three times the panel median (3.41), but with a welfare research base of 21.83 the L3 score of Kenya is 0.407, below the panel mean (0.475). Across the panel, the median AI share grew from 0.49\% in 2010 to 3.71\% in 2022 and the median precision agriculture research intensity from 0.54 to 3.41 works per 10,000, whereas the median welfare research base grew from 3.53 to 7.87. AI research thus grew more than sevenfold and precision agriculture research more than sixfold, against slightly more than a doubling of welfare research. These three variables measure research output, and the pathway from research output to a welfare harm is assumed (Table A1, Appendix). AI is also used to monitor welfare (Luo et al., 2023; Siachos et al., 2025). L3 therefore indicates where research capacity for automating animal agriculture grows fastest relative to the national capacity to evaluate that automation, and not where AI has harmed animals.

\subsection{Implications for Governance}

Our empirical research results in the following implications. First, national statistics do not currently support a cross-country measure of animal welfare enforcement. We were able to assemble 3,744 observed values on animal use, welfare legislation, and institutional conditions, and we were able to assemble nothing on inspection coverage, non-compliance rates, or enforcement actions on a common basis across 24 countries. \textit{Regulation (EU) 2017/625} requires member states to publish annual control reports, so that data exists for part of Europe, and its absence for the rest of the world is the single largest obstacle to measuring global welfare governance outcomes. Second, exposure and legal protection are conceptually different constructs. The WAP index grades legislation, whereas our AWPRI measures the number of animals exposed and the institutional conditions surrounding that exposure. The WAP index and our AWPRI do not correlate on either the overall grade ($\rho$ = $-$0.264, p = 0.213) or the farm animal grade ($\rho$ = $-$0.301, p = 0.153), indicating that legislative grade and exposure conditions provide complementary but not substitutable information. An accurate assessment of animal welfare governance requires both measures instead of either one. Third, where trade instruments analogous to the \textit{EU Deforestation Regulation} (European Parliament \& Council of the European Union, 2023) are extended to animal welfare, the compliance test has to attach to verifiable farm-level conduct. Nothing in this index, and nothing in the data sources we draw on, would support conditioning market access on a composite AWPRI score. Because the AWPRI has not been validated against inspection coverage, non-compliance rates, or other data on which welfare policy decisions are based, the AWPRI describes the relative positions of countries and is not a basis for allocating inspection effort, conditioning trade, or other policy decisions.

\section{Conclusions and Limitations}

Across the panel, research on automating animal agriculture grew faster than research on animal welfare over 2010--2022, and this research composition accounts for most of the upward trend in the AWPRI once the bounds are fixed across years (Sections 4.5 and 5.4). However, our index cannot indicate what such research endeavours do to any animal. The case for welfare-specific accountability mechanisms in regulatory frameworks for agricultural AI therefore rests on the evidence reviewed in Section 2.2 (Elliott \& Werkheiser, 2023; Tuyttens et al., 2022), and not on the AWPRI. Enforcement data represent the binding constraint on this field. Until inspection and non-compliance series are published on a common basis, no cross-country index can distinguish a country with strong statutes from a country that enforces them. An index of this kind is useful as a description of exposure and institutional conditions, not as a ranking of animal welfare outcomes.

It is noteworthy that our study is subject to the following limitations. First, the AWPRI measures exposure and governance capacity and does not measure animal welfare. No variable in the index records the state of any animal. The pathway from each indicator to a specified welfare harm is stated in Table A1 (Appendix), and for the three Layer 3 variables that pathway is assumed and not established. A reader who does not accept the amplification premise should read L1 and L2 on their own, and Section 4.8 reports what our global ranking looks like without L3. Second, layer-level equal weighting leads to unequal variable weights, since a Layer 3 variable carries 1/9 (= 1/3 x 1/3) of the composite against 1/15 (= 1/5 x 1/3) for a Layer 2 variable. Section 4.8 reports what happens under four alternative weightings and the rank correlation falls as low as 0.914, with one country moving nine places in our global ranking.

Third, min-max normalisation within year makes every score dependent on which other countries are in the panel. The leave-one-country rebuilds in Section 4.8 show that dependence is small within this panel, with 18 of the 24 rebuilds leaving every country's rank unchanged. Fourth, our panel holds 24 countries and is not globally representative. Countries that a scale-up version of the AWPRI project should include, namely Indonesia, Pakistan, and Ethiopia, are absent because we have not yet carried them through data collection, and their absence means that our study does not account for a large share of world livestock production. Fifth, our index contains no enforcement measure. Inspection coverage, non-compliance findings, and prosecution outcomes are not published on a common basis across included countries, and the civic space and civil liberties variables in L2 are conditions under which enforcement is more or less likely to occur. These variables cannot directly be served as enforcement measures. Sixth, our external test in Section 4.6 is against one instrument, the WAP's Animal Protection Index, which grades legislation and is itself an expert assessment and not an outcome measure. An outcome-based test against inspection coverage and non-compliance rates under \textit{Regulation (EU) 2017/625} would reach about nine of the 24 countries in this panel. We have not carried out that test in this research paper. Such a test would also need to account for differences between member states in the coverage and targeting of inspections. That test is the priority extension for the scale-up version of the AWPRI.

\textbf{Data availability:} The panel, the codebook, the robustness results, and every figure in this paper are published on our public dashboard at \url{https://awpri.aiinsocietyhub.com/}. The code we use to develop this public dashboard is stored in a public repository at \url{https://github.com/newlivehung123123/awpri}. This research paper is designed to present the methodology that the public dashboard builds upon and the empirical findings presented in the dashboard \textit{per se}.

\textbf{Ethics Statement:} This original research does not involve humans, animals, plants, biological material, protected or non-public datasets, collections or sites. Only secondary data from publicly accessible databases are extracted for statistical data analysis. Therefore, ethics approval is not required.

\section*{References}

\begingroup\raggedright\setlength{\leftskip}{0.4in}\setlength{\parindent}{-0.4in}\setlength{\parskip}{3pt}

Angrist, J. D., \& Pischke, J.-S. (2009). Mostly harmless econometrics: An empiricist's companion. Princeton University Press.

Animal Law Foundation. (2024). The enforcement problem: 2024 data. \url{https://animallawfoundation.org/enforcement}

Blattner, C. E., \& Bridges, J. C. (2024). For whose sake and benefit? A critical analysis of leading international treaty proposals to protect nonhuman animals. American Journal of Comparative Law, 72(1), 1--32. \url{https://doi.org/10.1093/ajcl/avae018}

Browning, H. (2022). Assessing measures of animal welfare. Biology \& Philosophy, 37(4), Article 36. \url{https://doi.org/10.1007/s10539-022-09862-1}

Callaway, B., \& Sant'Anna, P. H. C. (2021). Difference-in-differences with multiple time periods. Journal of Econometrics, 225(2), 200--230. \url{https://doi.org/10.1016/j.jeconom.2020.12.001}

Chaney, P., Jones, I. R., \& Narayan, N. (2024). Beyond the unitary state: Multi-level governance, politics, and cross-cultural perspectives on animal welfare. Animals, 14(1), Article 79. \url{https://doi.org/10.3390/ani14010079}

DataM Intelligence. (2024). AI in precision livestock farming market report 2024--2032. DataM Intelligence.

Elliott, K., \& Werkheiser, I. (2023). A framework for transparency in precision livestock farming. Animals, 13(21), Article 3358. \url{https://doi.org/10.3390/ani13213358}

European Parliament \& Council of the European Union. (2017). \textit{Regulation (EU) 2017/625} of the European Parliament and of the Council on official controls. Official Journal of the European Union. \url{http://data.europa.eu/eli/reg/2017/625/oj}

European Parliament \& Council of the European Union. (2023). \textit{Regulation (EU) 2023/1115} of the European Parliament and of the Council. Official Journal of the European Union. \url{http://data.europa.eu/eli/reg/2023/1115/oj}

Food and Agriculture Organization of the United Nations (FAO). (2023). FAOSTAT: Livestock primary data. \url{https://www.fao.org/faostat}

Food and Agriculture Organization of the United Nations (FAO). (2025). FAOLEX database, open data collection [Dataset]. FAO. \url{https://www.fao.org/faolex/en/}

Food and Agriculture Organization of the United Nations (FAO). (2026). FAOSTAT bulk downloads (Crops and livestock products, Food balance sheets, Macro-statistics, Population) [Dataset]. FAO. Retrieved August 4, 2026, from \url{https://bulks-faostat.fao.org/production}

H\aa{}rstad, R. M. B. (2024). The politics of animal welfare: A scoping review of farm animal welfare governance. Review of Policy Research, 41(4), 679--702. \url{https://doi.org/10.1111/ropr.12554}

Institute for Economics and Peace (IEP). (2024). Global peace index 2024. \url{https://www.economicsandpeace.org/wp-content/uploads/2024/06/GPI-2024-web.pdf}

Luo, S., Ma, Y., Jiang, F., Wang, H., Tong, Q., \& Wang, L. (2023). Dead laying hens detection using TIR-NIR-Depth images and deep learning on a commercial farm. \textit{Animals, 13}(11), Article 1861. \url{https://doi.org/10.3390/ani13111861}

Neethirajan, S. (2024). Artificial intelligence and sensor innovations: Enhancing livestock welfare with a human-centric approach. Human-Centric Intelligent Systems, 4(1), 77--92. \url{https://doi.org/10.1007/s44230-023-00050-2}

Organisation for Economic Co-operation and Development (OECD) \& Joint Research Centre (JRC). (2008). Handbook on constructing composite indicators: Methodology and user guide. OECD Publishing. \url{https://doi.org/10.1787/9789264043466-en}

Papakonstantinou, G. I., Voulgarakis, N., Terzidou, G., Fotos, L., Giamouri, E., \& Papatsiros, V. G. (2024). Precision livestock farming technology: Applications and challenges of animal welfare and climate change. Agriculture, 14(4), Article 620. \url{https://doi.org/10.3390/agriculture14040620}

Parlasca, M., Kn\"o\ss{}lsdorfer, I., Alemayehu, G., \& Doyle, R. (2023). How and why animal welfare concerns evolve in developing countries. Animal Frontiers, 13(1), 26--33. \url{https://doi.org/10.1093/af/vfac082}

Priem, J., Piwowar, H., \& Orr, R. (2022). OpenAlex: A fully-open index of scholarly works, authors, venues, institutions, and concepts. arXiv. \url{https://doi.org/10.48550/arXiv.2205.01833}

Schillings, J., Bennett, R., \& Rose, D. C. (2021). Exploring the potential of precision livestock farming technologies to help address farm animal welfare. \textit{Frontiers in Animal Science, 2}, Article 639678. \url{https://doi.org/10.3389/fanim.2021.639678}

Siachos, N., Griffiths, B. E., Wilson, J. P., Bedford, C., Anagnostopoulos, A., Neary, J. M., Smith, R. F., \& Oikonomou, G. (2025). Evaluation of a fully automated 2-dimensional imaging system for real-time cattle lameness detection using machine learning. \textit{Journal of Dairy Science, 108}(4), 4206--4224. \url{https://doi.org/10.3168/jds.2024-25940}

Tuyttens, F. A. M., Molento, C. F. M., \& Benaissa, S. (2022). Twelve threats of precision livestock farming (PLF) for animal welfare. Frontiers in Veterinary Science, 9, Article 889623. \url{https://doi.org/10.3389/fvets.2022.889623}

United Nations Development Programme (UNDP). (1990). Human development report 1990. \url{https://hdr.undp.org/system/files/documents/hdr1990encompletenostats.pdf}

V-Dem Institute. (2025). Country-year: V-Dem full + others version 15 [Dataset]. University of Gothenburg. \url{https://v-dem.net/data/the-v-dem-dataset/country-year-v-dem-fullothers-v15/}

Wolf, M. J., Emerson, J. W., Esty, D. C., de Sherbinin, A., \& Wendling, Z. A. (2022). 2022 environmental performance index. Yale Center for Environmental Law \& Policy.

World Animal Protection (WAP). (2020). Animal protection index 2020. \url{https://www.worldanimalprotection.us/siteassets/reports-programmatic/animal-protection-index-2020-report.pdf}

World Bank. (n.d.). World Bank country and lending groups. Retrieved August 2026, from \url{https://datahelpdesk.worldbank.org/knowledgebase/articles/906519}

Zhang, L., Guo, W., Lv, C., Guo, M., Yang, M., Fu, Q., \& Liu, X. (2024). Advancements in artificial intelligence technology for improving animal welfare: Current applications and research progress. Animal Research and One Health, 2(1), 93--109. \url{https://doi.org/10.1002/aro2.44}

\endgroup
\FloatBarrier
\section*{Appendix}
\setcounter{table}{0}\renewcommand{\thetable}{A\arabic{table}}\renewcommand{\theHtable}{A\arabic{table}}

\FloatBarrier
\begin{landscape}

\begingroup
\scriptsize
\setlength{\tabcolsep}{4pt}
\setlength{\LTpost}{3pt}
\begin{longtable}{@{}>{\raggedright\arraybackslash}p{\dimexpr 0.0720\linewidth-2\tabcolsep\relax}>{\raggedright\arraybackslash}p{\dimexpr 0.1057\linewidth-2\tabcolsep\relax}>{\raggedright\arraybackslash}p{\dimexpr 0.1746\linewidth-2\tabcolsep\relax}>{\raggedright\arraybackslash}p{\dimexpr 0.0873\linewidth-2\tabcolsep\relax}>{\raggedright\arraybackslash}p{\dimexpr 0.1470\linewidth-2\tabcolsep\relax}>{\raggedright\arraybackslash}p{\dimexpr 0.0842\linewidth-2\tabcolsep\relax}>{\raggedright\arraybackslash}p{\dimexpr 0.1118\linewidth-2\tabcolsep\relax}>{\raggedright\arraybackslash}p{\dimexpr 0.1164\linewidth-2\tabcolsep\relax}>{\raggedright\arraybackslash}p{\dimexpr 0.1011\linewidth-2\tabcolsep\relax}@{}}
\caption{AWPRI Constituent Variables, Sources, Welfare Pathway, and Coverage}\label{tab:A1}\\
\toprule
\textbf{Code} & \textbf{Variable} & \textbf{Source and query} & \textbf{Layer} & \textbf{Welfare pathway} & \textbf{Unit} & \textbf{Years Available} & \textbf{Transformation} & \textbf{Missing-Data Treatment} \\
\midrule
\endfirsthead
\multicolumn{9}{@{}l}{\normalsize\textbf{Table A1} (continued)}\\
\toprule
\textbf{Code} & \textbf{Variable} & \textbf{Source and query} & \textbf{Layer} & \textbf{Welfare pathway} & \textbf{Unit} & \textbf{Years Available} & \textbf{Transformation} & \textbf{Missing-Data Treatment} \\
\midrule
\endhead
\midrule
\multicolumn{9}{r@{}}{\textit{Continued on next page}}\\
\endfoot
\bottomrule
\endlastfoot
\textit{VAR\_01} & Farmed animals slaughtered per capita & FAOSTAT QCL (Crops and livestock products). Producing animals/slaughtered, summed over non-overlapping species series so one animal is counted once; FAO repeats the same head count on the meat, offal, fat and hides rows & L1, Animal-use exposure & Direct. Each slaughtered animal is an animal that lived under husbandry and died in an abattoir. The count is the number of animals exposed to whatever standards a country applies. & Slaughtered animals, per capita & 2010--2022 (13 years); all 24 countries & None before normalisation; min-max normalised within year to [0, 1] (Section 3.3); coded so that a higher value means more animals exposed, weaker welfare law, worse institutional conditions for enforcing welfare law, or research effort tilted further towards automating animal agriculture (Section 3.2). & No missing values by construction; the assembly raises an error on a missing value instead of filling one (Section 3.1). \\
\textit{VAR\_02} & Aquatic animal share of flesh supply & FAOSTAT Food Balance Sheets, item 2960 Fish, Seafood. Aquatic animal supply as a share of total flesh supply. FBS cannot separate farmed from wild-caught, so this is not an aquaculture share and is not named as one & L1, Animal-use exposure & Direct. Aquatic animals are excluded from most slaughter and transport welfare law, are killed in numbers that are counted by tonnage and not by individual, and their capacity for suffering is contested in a way that keeps them outside protective regimes. & Share of total flesh supply & Same & Same & Same \\
\textit{VAR\_05} & Meat supply per capita & FAOSTAT Food Balance Sheets, element 645. Meat supply quantity, kg per capita per year. FBS measures supply, not consumption, and the label says so & L1, Animal-use exposure & Direct. Volume of animal product supplied is the demand signal that sets how many animals are raised and how intensively. & kg per capita per year & Same & Same & Same \\
\textit{VAR\_07} & Animal share of protein supply & FAOSTAT Food Balance Sheets, element 674. Animal products (item 2941) as a share of animal plus vegetal (item 2903) protein supply, g/capita/day & L1, Animal-use exposure & Direct. How much of a population's protein comes from animals sets the scale of animal use per person independently of total food volume, and is the quantity a dietary transition would move. & Share of total protein supply & Same & Same & Same \\
\textit{VAR\_03} & Animal welfare legislation, stock & FAOLEX Open Data, complete collection. Binding instruments (Type of text Legislation or Regulation) carrying the FAO keyword `animal welfare', cumulative to the year, per 1,000 records indexed for that country at retrieval (4 August 2026), a denominator held fixed across years & L2, Governance capacity & Direct. A binding instrument is a rule about how animals may be kept, transported or killed. More of them means more conduct is regulated. The per-1,000 denominator is there because FAOLEX indexes some countries more than ten times as deeply as others (566 records for South Korea against 6,056 for the United Kingdom), and an unnormalised count measures the indexing. & Instruments per 1,000 indexed records, cumulative & Same & Min-max normalised within year to [0, 1], then inverted (1 $-$ \textit{v}), so that a larger stock of instruments is a lower score & Same \\
\textit{VAR\_04} & Rule of law & V-Dem v15, \textit{v2x\_\allowbreak rule}. Rule of law index, inverted so a high value is high risk & L2, Governance capacity & Indirect. Animal welfare law binds only where law binds. Where courts and inspectorates are weak, a statute on the books does not change what happens in a shed. & V-Dem index (0--1), inverted & Same & Inverted at source (1 $-$ \textit{v2x\_\allowbreak rule}), then min-max normalised within year to [0, 1]; not log-transformed & Same \\
\textit{VAR\_06} & Animal welfare legislation, five-year flow & FAOLEX Open Data, complete collection. Binding instruments carrying the keyword `animal welfare' dated within the preceding five years, per 1,000 indexed records & L2, Governance capacity & Direct. Recent instruments indicate a live legislative programme and not a body of law that stopped being added to. Stock and flow are separated because a country can have a large historic stock and no current activity. & Instruments per 1,000 indexed records, preceding five years & Same & Min-max normalised within year to [0, 1], then inverted (1 $-$ \textit{v}), so that a larger flow of instruments is a lower score & Same \\
\textit{VAR\_09} & Civic space & V-Dem v15, \textit{v2cseeorgs\_\allowbreak osp} and \textit{v2csprtcpt\_\allowbreak osp}. Mean of CSO entry and exit and CSO participatory environment, inverted & L2, Governance capacity & Indirect. Almost every documented case of intensive-farming cruelty in the historical record was surfaced by an NGO, a journalist or a whistleblower, not by an inspectorate. Where civil society cannot organise, harms are less likely to be identified at all. & Mean of two V-Dem indicators (0--4), each divided by 4, inverted & Same & Inverted at source (1 minus the mean of \textit{v2cseeorgs\_\allowbreak osp} and \textit{v2csprtcpt\_\allowbreak osp}, each divided by 4), then min-max normalised within year to [0, 1]; not log-transformed, because pooled skewness (2.06) stays above 1 after log(1 + \textit{x}) (1.72) & Same \\
\textit{VAR\_10} & Civil liberties & V-Dem v15, \textit{v2x\_\allowbreak civlib}. Civil liberties index, inverted & L2, Governance capacity & Indirect. Ag-gag statutes, restrictions on filming inside facilities and constraints on protest all operate through civil liberties, and each of them suppresses the evidence on which welfare enforcement depends. & V-Dem index (0--1), inverted & Same & Inverted at source (1 $-$ \textit{v2x\_\allowbreak civlib}), then min-max normalised within year to [0, 1]; not log-transformed, because pooled skewness (1.50) stays above 1 after log(1 + \textit{x}) (1.29) & Same \\
\textit{VAR\_12} & AI research intensity & OpenAlex. Works whose title or abstract contains `artificial intelligence', `machine learning', `deep learning', or `neural network', as a percentage of the country's total works, by publication year and author affiliation country; a work with authors in several countries counts once for each country & L3, AI amplification & Assumed. A large national AI research base is the capacity from which agricultural automation is built. That capacity reaching animal agriculture, and doing so in ways that worsen welfare and do not improve it, is a premise of this index and is not established by it. & Percentage of national works & Same & log(1 + \textit{x}) before normalisation (pooled skewness 1.98 before, 0.93 after), then min-max normalised within year to [0, 1] & Same \\
\textit{VAR\_13} & Precision agriculture research intensity & OpenAlex. Works whose title or abstract contains `precision agriculture', `PLF', `smart farming', or `digital agriculture', per 10,000 national works, counted as for VAR\_12 & L3, AI amplification & Assumed. Research on automating livestock production is the closest measurable proxy for technology that will be deployed on animals. Whether such technology harms or helps welfare depends on the application, and the direction assumed here is that automation at scale increases stocking density and reduces human contact. & Works per 10,000 national works & Same & log(1 + \textit{x}) before normalisation (pooled skewness 2.41 before, 0.51 after), then min-max normalised within year to [0, 1] & Same \\
\textit{VAR\_14} & Animal welfare research base & OpenAlex. Works whose title or abstract contains `animal welfare', `animal husbandry', `livestock welfare', or `farm animal', per 10,000 national works, counted as for VAR\_12. Plain animal welfare research, not AI-and-welfare, because a search for works containing both an AI term and a welfare term returned between 0 and 7 United Kingdom works a year, too few for an annual series & L3, AI amplification & Assumed. A country with a substantial welfare research base has expertise that can evaluate and contest new husbandry technology. The welfare research base sits in Layer 3 because the welfare research base is the quantity against which the other two Layer 3 variables are read, which is a claim about amplification. Placing the variable in Layer 2 instead returns a rank correlation of 0.930 against the index as built, with Argentina and Italy each moving six places. & Same & Same & log(1 + \textit{x}) before normalisation (pooled skewness 1.45 before, 0.17 after), then min-max normalised within year to [0, 1] and inverted (1 $-$ \textit{v}), so that a larger welfare research base is a lower score & Same \\
\textit{VAR\_08} & Public concern for animals & Hand-assigned constant. The variable did not vary within a country across 19 years and correlated with GDP per capita. No source was recorded. & Withdrawn & Not applicable & / & / & / & / \\
\textit{VAR\_11} & AI governance framework covering non-human stakeholders & The OECD.AI keyword search returned zero matches for all 25 countries across animal, welfare, sentience, non-human, livestock and species. That null is worth reporting and the paper reports it, but a constant carries no information about differences between countries. The published script recoded it to whether any national AI strategy exists, from a hand-typed dictionary, which measures technology policy maturity, and nothing about animals. & Withdrawn & Not applicable & / & / & / & / \\
\textit{VAR\_15} & Livestock AI patent intensity & No accessible source. EPO OPS and PATSTAT need credentials, Google Patents needs a billed BigQuery project, PATENTSCOPE has no API. The published series was a dictionary of 2022 counts divided by one 2022 GDP figure and scaled backwards at 20 per cent a year. \textit{VAR\_13} now carries the construct. & Withdrawn & Not applicable & / & / & / & / \\
\end{longtable}
\endgroup

\end{landscape}

\begin{table}[!htbp]
\centering
\caption{Observed Data Coverage by Country, All 24 Countries}
\label{tab:A2}
\small
\begin{tabular*}{\linewidth}{@{\extracolsep{\fill}}lllll@{}}
\toprule
\textbf{Country} & \textbf{Variables} & \textbf{Years} & \textbf{\textit{Values expected}} & \textbf{Values observed} \\
\midrule
Argentina & 12 & 2010--2022 (13) & 156 & 156 \\
Australia & 12 & 2010--2022 (13) & 156 & 156 \\
Brazil & 12 & 2010--2022 (13) & 156 & 156 \\
Canada & 12 & 2010--2022 (13) & 156 & 156 \\
China & 12 & 2010--2022 (13) & 156 & 156 \\
Denmark & 12 & 2010--2022 (13) & 156 & 156 \\
France & 12 & 2010--2022 (13) & 156 & 156 \\
Germany & 12 & 2010--2022 (13) & 156 & 156 \\
India & 12 & 2010--2022 (13) & 156 & 156 \\
Italy & 12 & 2010--2022 (13) & 156 & 156 \\
Kenya & 12 & 2010--2022 (13) & 156 & 156 \\
Mexico & 12 & 2010--2022 (13) & 156 & 156 \\
Netherlands & 12 & 2010--2022 (13) & 156 & 156 \\
New Zealand & 12 & 2010--2022 (13) & 156 & 156 \\
Nigeria & 12 & 2010--2022 (13) & 156 & 156 \\
Poland & 12 & 2010--2022 (13) & 156 & 156 \\
South Africa & 12 & 2010--2022 (13) & 156 & 156 \\
South Korea & 12 & 2010--2022 (13) & 156 & 156 \\
Spain & 12 & 2010--2022 (13) & 156 & 156 \\
Sweden & 12 & 2010--2022 (13) & 156 & 156 \\
Thailand & 12 & 2010--2022 (13) & 156 & 156 \\
United Kingdom & 12 & 2010--2022 (13) & 156 & 156 \\
United States & 12 & 2010--2022 (13) & 156 & 156 \\
Vietnam & 12 & 2010--2022 (13) & 156 & 156 \\
\bottomrule
\end{tabular*}
\par\smallskip
{\footnotesize\raggedright \textit{Note. Each country contributes 12 variables $\times$ 13 years = 156 values. Observed means the value is read from the named public source for that country and year. The panel contains no imputed, interpolated, carried-forward, or reconstructed values, and the assembly raises an error on a missing value instead of filling it.}\par}
\end{table}

\begingroup
\footnotesize
\setlength{\tabcolsep}{3pt}
\setlength{\LTpost}{3pt}
\begin{longtable}{@{}>{\raggedright\arraybackslash}p{\dimexpr 0.2900\linewidth-2\tabcolsep\relax}>{\raggedright\arraybackslash}p{\dimexpr 0.1200\linewidth-2\tabcolsep\relax}>{\raggedright\arraybackslash}p{\dimexpr 0.0800\linewidth-2\tabcolsep\relax}>{\raggedright\arraybackslash}p{\dimexpr 0.0900\linewidth-2\tabcolsep\relax}>{\raggedright\arraybackslash}p{\dimexpr 0.1100\linewidth-2\tabcolsep\relax}>{\raggedright\arraybackslash}p{\dimexpr 0.3100\linewidth-2\tabcolsep\relax}@{}}
\caption{Robustness of the Construction, 46 Rebuilds of the Index}\label{tab:A3}\\
\toprule
\textbf{Rebuild} & \textbf{Spearman $\rho$ vs index as built} & \textbf{Max rank move} & \textbf{\textit{Mean |rank move|}} & \textbf{\textit{n countries}} & \textbf{Largest movers} \\
\midrule
\endfirsthead
\multicolumn{6}{@{}l}{\normalsize\textbf{Table A3} (continued)}\\
\toprule
\textbf{Rebuild} & \textbf{Spearman $\rho$ vs index as built} & \textbf{Max rank move} & \textbf{\textit{Mean |rank move|}} & \textbf{\textit{n countries}} & \textbf{Largest movers} \\
\midrule
\endhead
\midrule
\multicolumn{6}{r@{}}{\textit{Continued on next page}}\\
\endfoot
\bottomrule
\multicolumn{6}{@{}p{\dimexpr\linewidth-2\tabcolsep\relax}@{}}{\footnotesize\raggedright\rule{0pt}{3ex}\textit{Note. Each row rebuilds the index under one alternative construction and compares the resulting 2022 country ordering to the index as built. Max move is the largest number of rank places any single country changes. Leave-one-country rebuilds compare the remaining 23 countries only. No threshold is asserted at which the construction becomes robust.}}\\
\endlastfoot
z-score within year & 0.9939 & 2 & 0.42 & 24 & NG~$-$2, FR~+2, AR~$-$1, DE~+1 \\
min-max on pooled bounds, all years at once & 0.9817 & 3 & 0.83 & 24 & AR~+3, IT~$-$3, CA~$-$2, NZ~+2 \\
z-score on pooled bounds & 0.9852 & 4 & 0.67 & 24 & NG~$-$4, AR~+3, FR~+1, NL~$-$1 \\
layer weights from PC1 of the layer scores & 0.9948 & 2 & 0.42 & 24 & MX~+2, CA~$-$1, DE~+1, DK~$-$1 \\
layer weights 2:1:1, exposure doubled & 0.9139 & 9 & 2.00 & 24 & NG~$-$9, IN~$-$6, NL~+4, ZA~$-$3 \\
layer weights 1:2:1, governance doubled & 0.9261 & 5 & 1.92 & 24 & AR~+5, NG~+5, KE~+5, NL~$-$4 \\
layer weights 1:1:2, AI doubled & 0.9600 & 4 & 1.50 & 24 & BR~$-$4, IT~+4, ES~+3, PL~$-$3 \\
without \textit{VAR\_01} \textit{farmed\_\allowbreak animals\_\allowbreak per\_\allowbreak capita} & 0.9591 & 4 & 1.42 & 24 & PL~$-$4, ES~+4, IT~+3, BR~$-$3 \\
without \textit{VAR\_02} \textit{aquatic\_\allowbreak share\_\allowbreak of\_\allowbreak flesh} & 0.9139 & 9 & 1.92 & 24 & NG~$-$9, IN~$-$7, AR~+4, US~+3 \\
without \textit{VAR\_05} \textit{meat\_\allowbreak supply\_\allowbreak kg} & 0.9843 & 3 & 0.75 & 24 & NG~+3, AR~$-$2, DK~$-$2, NL~+2 \\
without \textit{VAR\_07} \textit{animal\_\allowbreak protein\_\allowbreak share} & 0.9791 & 5 & 0.83 & 24 & NG~+5, KE~+2, NZ~+2, CA~$-$2 \\
without \textit{VAR\_03} \textit{welfare\_\allowbreak legislation\_\allowbreak stock\_\allowbreak risk} & 0.9896 & 4 & 0.50 & 24 & MX~+4, AR~$-$1, ES~$-$1, US~$-$1 \\
without \textit{VAR\_04} \textit{rule\_\allowbreak of\_\allowbreak law\_\allowbreak risk} & 0.9809 & 4 & 0.92 & 24 & NG~$-$4, FR~+3, ZA~+2, BR~$-$2 \\
without \textit{VAR\_06} \textit{welfare\_\allowbreak legislation\_\allowbreak flow\_\allowbreak risk} & 0.9904 & 3 & 0.58 & 24 & MX~+3, US~$-$2, KE~$-$1, AU~$-$1 \\
without \textit{VAR\_09} \textit{civic\_\allowbreak space\_\allowbreak risk} & 0.9878 & 3 & 0.67 & 24 & AR~+3, FR~$-$2, PL~$-$2, NL~$-$2 \\
without \textit{VAR\_10} \textit{civil\_\allowbreak liberties\_\allowbreak risk} & 0.9887 & 3 & 0.58 & 24 & AR~+3, NG~$-$2, IT~$-$2, NZ~+2 \\
without \textit{VAR\_12} \textit{ai\_\allowbreak research\_\allowbreak intensity} & 0.9478 & 6 & 1.50 & 24 & AR~+6, KE~+5, AU~$-$4, IT~$-$3 \\
without \textit{VAR\_13} \textit{precision\_\allowbreak ag\_\allowbreak intensity} & 0.9643 & 4 & 1.50 & 24 & FR~+4, IT~$-$4, DE~+3, GB~+3 \\
without \textit{VAR\_14} \textit{animal\_\allowbreak welfare\_\allowbreak research\_\allowbreak risk} & 0.8652 & 10 & 2.42 & 24 & US~$-$10, AR~$-$6, IT~+6, KE~+6 \\
without L1 entirely & 0.7504 & 15 & 3.50 & 24 & KE~+15, NG~+8, ZA~+8, US~$-$7 \\
without L2 entirely & 0.6809 & 10 & 4.67 & 24 & NG~$-$10, ES~+9, MX~$-$8, CN~$-$8 \\
without L3 entirely & 0.8400 & 9 & 2.67 & 24 & NZ~+9, IN~$-$9, AR~+8, IT~$-$7 \\
leave out AU (Australia) & 1.0000 & 0 & 0.00 & 23 & none \\
leave out BR (Brazil) & 1.0000 & 0 & 0.00 & 23 & none \\
leave out CA (Canada) & 1.0000 & 0 & 0.00 & 23 & none \\
leave out FR (France) & 1.0000 & 0 & 0.00 & 23 & none \\
leave out DE (Germany) & 1.0000 & 0 & 0.00 & 23 & none \\
leave out IN (India) & 1.0000 & 0 & 0.00 & 23 & none \\
leave out IT (Italy) & 1.0000 & 0 & 0.00 & 23 & none \\
leave out NL (Netherlands) & 1.0000 & 0 & 0.00 & 23 & none \\
leave out NZ (New Zealand) & 0.9980 & 1 & 0.17 & 23 & IT~$-$1, DE~+1, DK~$-$1 \\
leave out KR (South Korea) & 1.0000 & 0 & 0.00 & 23 & none \\
leave out ES (Spain) & 1.0000 & 0 & 0.00 & 23 & none \\
leave out SE (Sweden) & 0.9990 & 1 & 0.09 & 23 & IT~$-$1, FR~+1 \\
leave out GB (United Kingdom) & 1.0000 & 0 & 0.00 & 23 & none \\
leave out US (United States) & 0.9960 & 2 & 0.26 & 23 & FR~+2, IT~$-$1, DE~+1 \\
leave out AR (Argentina) & 1.0000 & 0 & 0.00 & 23 & none \\
leave out CN (China) & 0.9941 & 2 & 0.43 & 23 & MX~+2, BR~+1, CA~$-$1 \\
leave out DK (Denmark) & 1.0000 & 0 & 0.00 & 23 & none \\
leave out KE (Kenya) & 1.0000 & 0 & 0.00 & 23 & none \\
leave out MX (Mexico) & 1.0000 & 0 & 0.00 & 23 & none \\
leave out NG (Nigeria) & 0.9980 & 1 & 0.17 & 23 & IT~$-$1, CA~$-$1, FR~+1 \\
leave out PL (Poland) & 0.9951 & 2 & 0.26 & 23 & FR~+2, NG~$-$2, CA~$-$1 \\
leave out ZA (South Africa) & 1.0000 & 0 & 0.00 & 23 & none \\
leave out TH (Thailand) & 1.0000 & 0 & 0.00 & 23 & none \\
leave out VN (Vietnam) & 1.0000 & 0 & 0.00 & 23 & none \\
\end{longtable}
\endgroup

\clearpage
\section*{Supplementary Material(s)}
\setcounter{table}{0}\renewcommand{\thetable}{S\arabic{table}}\renewcommand{\theHtable}{S\arabic{table}}

\begin{table}[!htbp]
\centering
\caption{AWPRI Trend Projections to 2030 (95\% Prediction Intervals)}
\label{tab:S1}
\small
\begin{tabular*}{\linewidth}{@{\extracolsep{\fill}}llllll@{}}
\toprule
\textbf{Country} & \textbf{AWPRI 2022} & \textbf{AWPRI 2030} & \textbf{95\% PI lower} & \textbf{95\% PI upper} & \textbf{Slope p} \\
\midrule
China & 0.7027 & 0.7357 & 0.6825 & 0.7889 & 0.054 ns \\
Vietnam & 0.6631 & 0.7354 & 0.5675 & 0.9033 & 0.038 * \\
Thailand & 0.6116 & 0.6439 & 0.5091 & 0.7786 & 0.304 ns \\
India & 0.5780 & 0.6346 & 0.5778 & 0.6914 & < 0.001 *** \\
South Korea & 0.5453 & 0.6553 & 0.5617 & 0.7488 & < 0.001 *** \\
United States & 0.5045 & 0.5517 & 0.4883 & 0.6151 & 0.026 * \\
Brazil & 0.4919 & 0.4730 & 0.4056 & 0.5405 & 0.616 ns \\
Canada & 0.4805 & 0.4636 & 0.3853 & 0.5419 & 0.430 ns \\
Australia & 0.4788 & 0.4851 & 0.4274 & 0.5428 & 0.353 ns \\
Mexico & 0.4741 & 0.4713 & 0.4181 & 0.5246 & 0.807 ns \\
Spain & 0.4629 & 0.4420 & 0.3726 & 0.5113 & 0.518 ns \\
Poland & 0.4520 & 0.4718 & 0.4051 & 0.5385 & 0.001 ** \\
Nigeria & 0.4376 & 0.4025 & 0.2736 & 0.5313 & 0.686 ns \\
Italy & 0.4354 & 0.4989 & 0.4148 & 0.5831 & 0.005 ** \\
France & 0.4351 & 0.4636 & 0.4005 & 0.5266 & 0.025 * \\
Argentina & 0.4270 & 0.3912 & 0.2733 & 0.5090 & 0.209 ns \\
South Africa & 0.4204 & 0.3757 & 0.2881 & 0.4634 & 0.935 ns \\
Netherlands & 0.3932 & 0.4332 & 0.3648 & 0.5015 & 0.048 * \\
United Kingdom & 0.3843 & 0.4210 & 0.3959 & 0.4462 & < 0.001 *** \\
New Zealand & 0.3784 & 0.3948 & 0.2669 & 0.5226 & 0.882 ns \\
Denmark & 0.3640 & 0.3886 & 0.3121 & 0.4651 & 0.023 * \\
Germany & 0.3617 & 0.3532 & 0.3166 & 0.3898 & 0.638 ns \\
Kenya & 0.3386 & 0.3014 & 0.1787 & 0.4242 & 0.729 ns \\
Sweden & 0.3161 & 0.3292 & 0.2594 & 0.3991 & 0.064 ns \\
\bottomrule
\end{tabular*}
\par\smallskip
{\footnotesize\raggedright \textit{Note. Projections are OLS linear trend extrapolations fitted on 2010--2022 and are not forecasts from a fitted time-series model. Slope p is the p value of the trend coefficient. Where slope p exceeds 0.05, the projection extrapolates a slope the data do not distinguish from zero and should be read as such. ns = non-significant.}\par}
\end{table}

\end{document}